\documentclass[aps,pre,twocolumn,superscriptaddress,superscriptreference]{revtex4-1}

\usepackage{amsmath,bbold,bm,amssymb,scalerel,mathtools}
\usepackage{graphicx}
\usepackage{color}
\usepackage{enumitem}
\usepackage{algorithm,algpseudocode}
\usepackage{multirow}
\usepackage{colortbl,booktabs}
\usepackage{placeins}
\usepackage[usenames,dvipsnames]{xcolor}
\usepackage[colorlinks, linkcolor=BrickRed, urlcolor=blue!50!black, citecolor=blue!50!black]{hyperref}
\usepackage{float}

\newcommand*{\citen}{}
\DeclareRobustCommand*{\citen}[1]{%
	\begingroup
	\romannumeral-`\x 
	\setcitestyle{numbers}%
	\cite{#1}%
	\endgroup
}

\usepackage[normalem]{ulem}

\newcommand\revision[1]{\textcolor{black}{#1}}

\newcommand{\gj}[1]{\textcolor{black}{#1}}
\newcommand{\fs}[1]{\textcolor{black}{#1}}

\begin{document}

\title{Stability of branched tubular membrane structures}

\author{Maike Jung}
\email{maike\_lauf@web.de}
\affiliation{Institut f\"ur Physik, Johannes Gutenberg-Universit\"at Mainz, 
	Staudingerweg 9, 55128 Mainz, Germany}

\author{Gerhard Jung}
	\affiliation{Laboratoire Charles Coulomb (L2C), Université de Montpellier, CNRS, 34095 Montpellier, France}


\author{Friederike Schmid}
\email{friederike.schmid@uni-mainz.de}
\affiliation{Institut f\"ur Physik, Johannes Gutenberg-Universit\"at Mainz, 
	Staudingerweg 9, 55128 Mainz, Germany}

\begin{abstract}

We study the energetics and stability of branched tubular membrane
structures by computer simulations of a triangulated network model.
We find that triple (Y-)junctions can be created and stabilized by
applying mechanical forces, if the angle between branches is $120^o$.
The same holds for tetrahedral junctions with tetraeder angles.
If the wrong angles are enforced, the branches coalesce to {a linear structure, a pure tube}.  After releasing the 
mechanical force, Y-branched structures remain metastable if one 
constrains the enclosed volume and the {average} curvature 
(the area difference) to a fixed value; tetrahedral junctions however 
split up into two Y-junctions. 
Somewhat counterintuitively, the \revision{energy cost} of \revision{adding a} Y-branch is negative in structures with fixed 
\revision{surface area and} tube diameter,  even if one accounts for the positive contribution  of the  additional branch end. For fixed {average curvature},  however, adding a  branch also enforces a thinning of tubes,  therefore the  overall  curvature energy cost is positive. Possible implications for the stability of branched 
networks structures in cells are discussed.

\end{abstract}

\maketitle

%
%

Tubular membrane network structures are abundant in biological
cells, for example in the Golgi complex \citen{Mollenhauer1998,
MatteisGolgi2008} and the endoplasmic reticulum \citen{WestrateER2015,
Lue2020}.  Such tubular networks are highly dynamic structures
\cite{Powers2017}, in which new tubes are constantly 
created and existing tubes are merged or dissolved.  
Potential physiological roles of the three-dimensional tubular network 
spanning the endoplasmic
reticulum include membrane trafficking, lipid metabolism and
autophagy, i.e. the cleaning mechanism of the cell \cite{Lue2020}.
The function of the tubular network in the Golgi apparatus appears to
be the interconnection of different building blocks, which can also
induce structural rearrangements during cell differentiation
\cite{Saraste2019}.  Membrane nanotubes have also been found to
generally enhance intercellular transport \cite{Sowinski2008}.
Understanding the formation and stability of tubular networks is thus
a critical problem in the fields of biology, biophysics and soft
matter.

The formation of tubular structures and membrane networks can be
induced by various different mechanisms, which can be classified into
different categories \cite{Roux2013}.  The most obvious way of
creating tubular structures is by a force acting on a localized point
on the membrane surface. This force can be induced by growing
filaments (filament bundles) which are attached to the membrane
\cite{Miyata1992,Miyata1999} or by a concerted action of molecular
motors \cite{DerenyiProst2002,Koster2003,Leduc2004,Campas2008,Nambiar2009,
Du2016}.  
Other mechanisms for tube formation include scaffolding, in which
proteins are polymerizing on the surface of the membrane, effectively
forcing the membrane to adopt the shape of the proteins
\cite{Footer2007,Roux2010}, and the adsorption or inclusion of
curvature-inducing proteins, which have been widely observed in nature
\cite{Tsafrir2003,Shibata2009} and can induce either positive or
negative curvature \cite{Campelo2008,Frost2009}. 
For example, reticulon has been found to induce the tubular network
structure in the endoplasmic reticulum
\citen{Voeltz2006,Hu2008,ShemeshER2014}.  

From a theoretical point of view, membrane shapes have been studied
intensely for many decades \cite{SeifertLipowsky1995,
ReviewRamakrishnan2018, ReviewLipowsky2021}, often using elastic
continuum models based on the  Canham-Helfrich theory
\cite{Canham1970,Helfrich1973,Evans1974}.  Already for structures with
simple sphere topology, the shape diagrams were found to be
surprisingly complex, with first and second order transitions between
prolate, oblate, pear and stomatocyte shapes
\cite{SeifertLipowsky1991,SeifertLipowsky1995,Gompper1995,Vanhille-Campos2021}.  
%
%
The process of mechanically pulling tubes from vesicles has been
investigated in detail by experiment, theory and
simulation \cite{Bo1989, Bukman1996, Calladine2002, Seifert2004,
Koster2005, Golushko2016, Noguchi2021,Paraschif2021} and found to be 
accompanied by a free energy barrier \cite{Koster2005}, suggesting that 
it might be possible to create metastable tubular structures using 
mechanical forces (e.g., molecular motors). Indeed, Bahrami et
al \cite{TriangulationBahrami2017} have recently demonstrated by
computer simulations that {linear tubular structures} can be 
metastable even in the absence of forces and curvature-inducing 
proteins, as long as the enclosed volume is kept fixed. This is 
due to the existence of a free energy barrier between the 
{linear} tube shape and the true minimum-energy shapes, 
which are oblate and prolate structures for thick tubes and
stromatocytes in the case of thin tubes.

While the (meta)stability of {linear tubular structures} has 
been analyzed in some detail, a network has a second fundamental 
building block, i.e. the junctions where several tubes merge. 
Detailed theoretical analyses of such branched structures, 
comparable to the ones for cylindrical tubes, however, are 
still missing. In the present Letter, we aim to fill this gap. 
We will first consider force-stabilized branched structures 
and examine their stability. Then we will establish conditions under
which force-free branched structures can be metastable.


{\em Model and method.} 
Our starting point is the simplest continuum description of two-dimensional fluid membranes on large scales, the so-called
Helfrich Hamiltonian \cite{Canham1970,Helfrich1973,Evans1974}. 
\begin{equation}
\label{eq:HelfrichHamiltonian}
H_\text{cv} = \frac{\kappa}{2}\int \text{d}A \: K^2
     + \bar{\kappa} \int \text{d}A K_{\text{G}}.
\end{equation}
Here $\kappa$, $\bar{\kappa}$ are curvature moduli (for lipid
membranes, $\kappa$ is typically of order $20 k_B T$
\cite{RAWICZ2000}), $K$ is the total curvature, and $K_{\text{G}}$
the Gaussian curvature.  We consider closed structures with fixed
sphere topology, hence the last term is a constant according to the
Gauss-Bonnet theorem \citen{DoCarmoDiffGeo1976} and can be omitted. We
note that we have not included a spontaneous curvature term in Eq.\
(\ref{eq:HelfrichHamiltonian}). Instead, we will discuss the effect of
imposing an integrated average curvature $\int \text{d}A \: K$ in the
spirit of the area difference elasticity (ADE) model
\cite{ADE_Bozic1992,ADE_Wiese1992,ADE_Heinrich_1993,ADE_Miao_1994}.
The physical origin of this global curvature could be asymmetric {numbers of lipid in the inner and outer membrane leaflet} 
(``area difference'')
\cite{BC_1974,Evans1974,BC_1989,ADE_Bozic1992,ADE_Wiese1992,
ADE_Heinrich_1993,ADE_Miao_1994,Ziherl2005}. 

The theory is solved numerically using a dynamically-triangulated
surface model \cite{
TriangulationItzykson1986,
TriangulationKantor1987,
TriangulationGompper1996,
TriangulationGompper1997,
TriangulationJulicher1996, TriangulationSaric2012,
TriangulationBahrami2012, TriangulationRamakrishnan2013,
TriangulationBahrami2017, TriangulationVahid2017, TriangulatedLi2018,
TriangulationHoore2018, BIAN2020}. Specifically, we use the version 
of Noguchi and Gompper \cite{TriangulationNoguchi2004} which is
described in detail in Ref.\ \cite{TriangulationNoguchi2005}.  The
surface is described by a network of $N$ vertices that are connected
by bonds in a triangular network structure ($N_\Delta= 2(N-2)$
triangles), and the simulation is a combination of Brownian dynamics
(node motion) and Monte Carlo moves (bond flips). 
\revision{We fix the area ($A = A_0$) and in some simulations
also the enclosed volume $V$ and the dimensionless average curvature (the area difference) \cite{ADE_Bozic1992,ADE_Wiese1992,ADE_Heinrich_1993,ADE_Miao_1994, TriangulationBahrami2017} 
$\Delta a = \frac{1}{4\sqrt{\pi A_0}} \int \text{d}A \: K$ by 
introducing harmonic constraint potentials with spring constants
$k_{\text{A}}, k_{\text{V}}$, and $k_{\Delta a}$.}
Details of the
implementation can be found in Supplementary Information (SI). 

In the following, results are given in units of $l_b$ (typical 
bond length), $\epsilon=\frac{\kappa}{20}$ (energy unit)
and $\tau = \revision{l_b} \sqrt{m \epsilon^{-1}}$ (time unit),
where $ m $ is the mass of the vertices.  Unless stated otherwise, the
remaining parameters  are $k_{\text{B}} T = 1\revision{\epsilon}$,  $N=2562$, 
$A_0 = 0.41 \revision{l_b^2 \cdot} N_\Delta$, $k_{\text{A}}=2 \revision{\epsilon/l_b^{2}}$, \revision{$k_{\text{V}}=k_{\Delta a} = 0$}, and the simulation time step is 
$\Delta t = 10^{-4} \revision{\tau}$. 
\revision{Constraints on $V$ and/or $\Delta a$ are imposed by 
setting $k_{\text{V}}=1 \revision{\epsilon/l_b^3}$ and/or 
$k_{\Delta a} = 1\epsilon$}. The enclosed
volume will be characterized by the dimensionless quantity $\nu=6
\sqrt{\pi/A^3} V$. The reference values of $\nu$ and $\Delta a$ for
perfect spheres are thus \mbox{$\nu = \Delta a = 1$}.

 
\begin{figure}	
		\centering
		\includegraphics[width=1.0\linewidth]
        {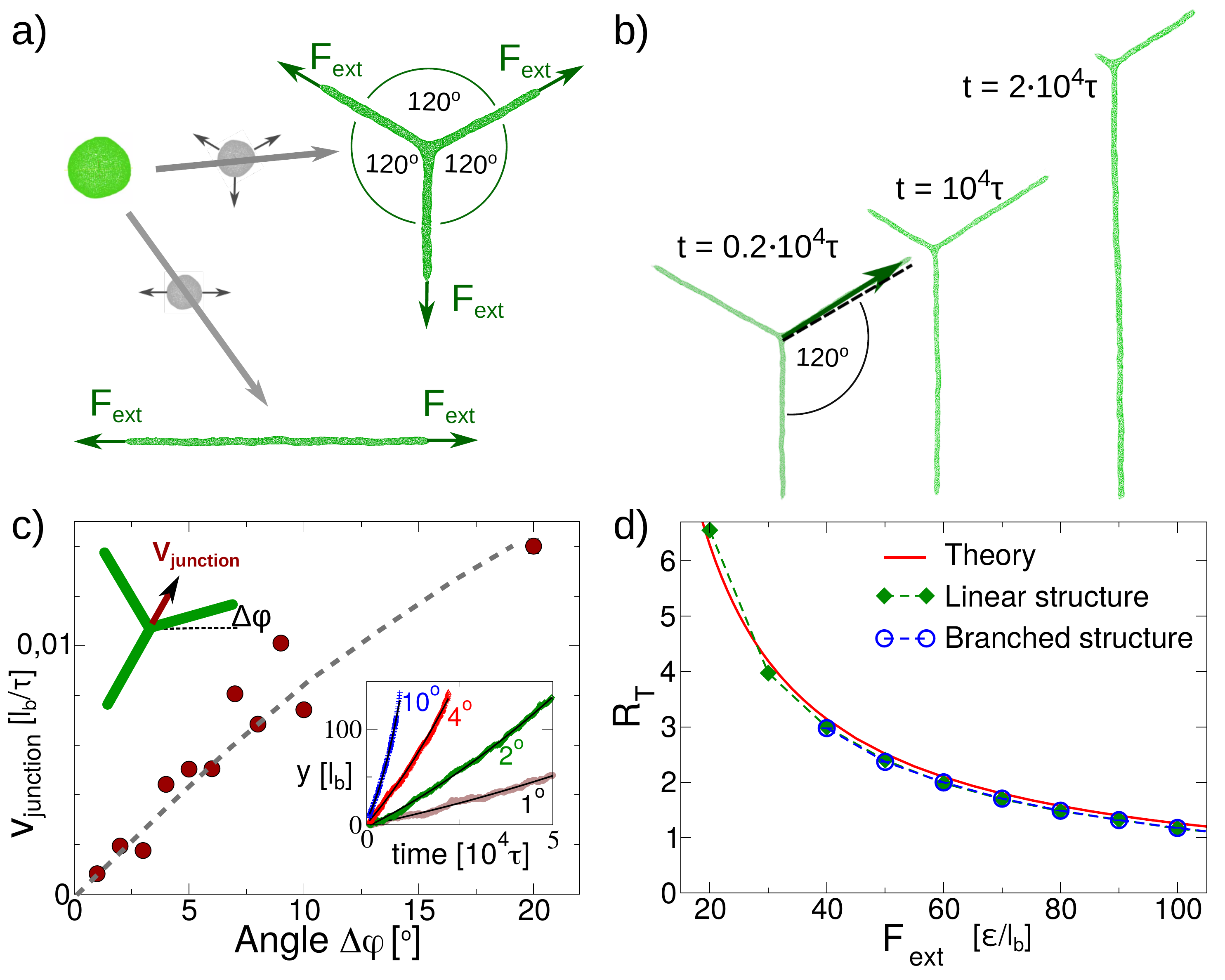}
	\caption{{Force-stabilized tubular structures}.
    {(a) Illustration of creation process.
    The starting point is a force-free spherical vesicle. To create
    linear structures, two opposing forces are applied at
    opposing vertices while keeping the area $A$ fixed 
    (no other constraints). Y-branched structures are obtained
    by applying three co-planar forces with angles 120${}^o$
    to each other.}
    (b) Evolution of a {Y-branched structure with time t}
     {at $F_{\text{ext}}=90\revision{\epsilon/l_b}$} if the
     direction of one {applied} force deviates from the symmetric
     direction by $\Delta \varphi = 4^o$ \cite{note}.
    (c) Initial velocity of the junction as a function of 
     $\Delta \varphi$ {at $F_{\text{ext}}=90\revision{\epsilon/l_B}$}.
     Dashed line is a guide for the eye. Inset shows the 
     \revision{displacement}
     $y$ of the junction \revision{from its initial position}  
     versus time for different $\Delta \varphi$ 
     as indicated, along with a quadratic fit to
     $y=y_0 + v_{\text{junction}} t + b t^2$ (black lines).
    (d) Tube radius versus applied force $F_{\text{ext}}$ 
     for {linear} (green diamond) and branched structures 
     (blue circles), compared with theory (red line).
	\label{fig:fig1}
    }
\end{figure}

{\em Force-stabilized {linear} and branched {tubular} structures.} 
To create {tubular} structures,  
forces with amplitude $F_{\text{ext}}$ are applied to a set of $n$   
vertices such that the total force is zero ($n=2,3,4$). For $n=2$, 
{linear  tubes} are obtained. For $n=3$, a branched structure with a
Y-junction  can be stabilized, provided the forces {lie in one plane
and} have an angle of $120^\circ$  to each other 
(see Fig.\ \ref{fig:fig1}a), otherwise one creates 
{linear structures} as well. {Using $n=4$, one can create 
mechanically forced tetrahedral junctions; all other four-fold junctions
are unstable and separate into Y-junctions}
(see SI, Fig.\ 1 {and movies {\tt 4fold.mp4}, {\tt 4fold\_twisted.mp4}, {\tt tetrahedral.mp4}}).

Y-junctions with fixed angle $120^\circ$ are characteristic of the
so-called Fermat point, the state that minimizes the total tube length
of a network if the tube ends are kept at fixed positions.  In
experimental studies, artificial surfactant and liposome networks with
fixed tube ends were found to always evolves towards the Fermat-point
\cite{Lobovkina2004, Lobovkina2006, Lobovkina2008}.  Our simulations
show that these $120^o$ Y-junctions remain the only stable triple
junctions even in situations where the tube ends are mobile.
Fig.\ \ref{fig:fig1}b) shows the effect of slightly perturbing the
angle of one applied force from $120^\circ$, starting from the
configuration \ref{fig:fig1}a): The junction starts moving in the
direction of the smallest angle until it disappears, with a velocity
that is roughly proportional to the distortion $\Delta \varphi$ (Fig.\
\ref{fig:fig1}c). 

For stable branched structures, the presence of the
junction has little effect on the structure of the connected tubes.
The tube radius $R_T$ as a function of the applied force $F_{\text{ext}}$ is
the same for {linear and branched structures} and consistent with the 
theoretical estimate \cite{Bo1989,DerenyiProst2002} $ R_{\text{T}} 
= {2 \pi \kappa}/{F_{\text{ext}}}$.  (Fig.\ \ref{fig:fig1}d).

\begin{figure}[t]
		\centering
		\includegraphics[width=1.0\linewidth]
        {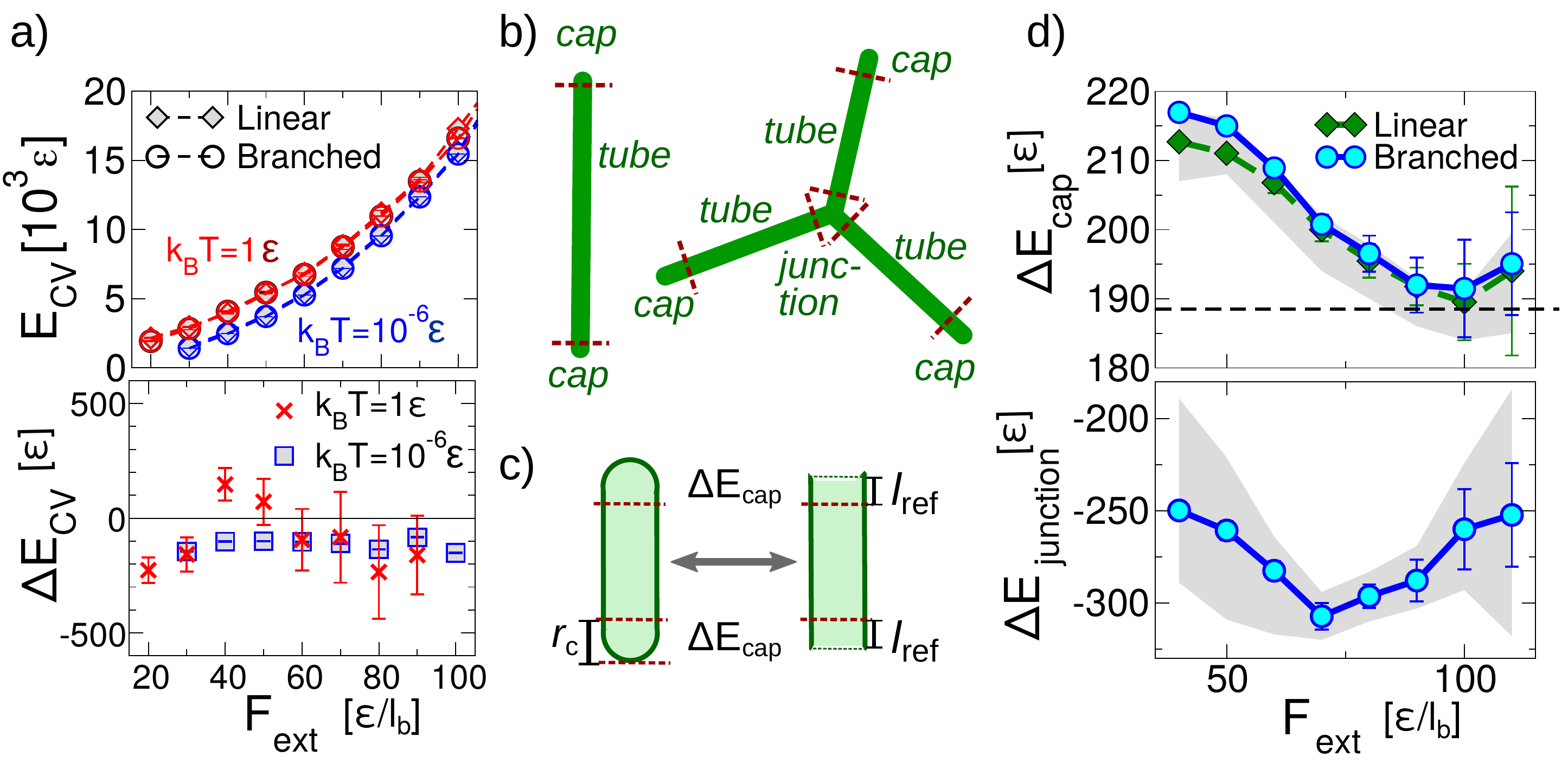}
	\caption{{Curvature energies of force-stabilized tubular structures}.
    (a) Top: Elastic energy for {linear} (diamonds) and branched
    (circles) structures as a function of applied force
    $F_{\text{ext}}$ at $k_B T= 1 \revision{\epsilon}$ (red) and
    $k_B T = 10^{-6} \revision{\epsilon} $ (blue). 
    Bottom: {Difference between the curvature energy of
    \revision{branched and linear} structures}
    $k_B T = 1 \revision{\epsilon}$ (red crosses) and 
    $k_B T = 10^{-6} \revision{\epsilon}$ (blue squares).
    (b) Cartoon showing dissection of structures into
    {tubes}, caps and junctions (see text).
    (c) Cartoon illustrating the definition of excess energies:
    The energy of a tubular structure is compared to that of a
    reference tubular section with the same area.
    (d) Excess curvature energy of caps (top) and junctions (bottom)
    for {linear} (green diamonds) and branched (blue circles)
    structures, obtained at $k_B T = 10^{-6} \revision{\epsilon}$. Dashed line
    (top) shows theoretical value for ideal semispherical caps.
    Symbols/lines show values obtained with cutoff parameters
    $r_{c,\text{cap}}=8 \revision{l_b}$ and $r_{c,\text{junction}}=20 \revision{l_b}$. 
    {Grey shaded areas indicate spread of results if one 
    varies} the cutoff between $r_{c,\text{cap}} \in [7,10] \revision{l_b}$ 
    and $r_{c,\text{junction}} \in [10,25] \revision{l_b}$.
	\label{fig:fig2}
    }
\end{figure}

Next we analyze the curvature energy (\ref{eq:HelfrichHamiltonian}) of
the different structures. Fig.\ \ref{fig:fig2}a) shows the results at
temperature $k_BT = 1$ and after annealing to $k_B T = 10^{-6} 
\revision{\epsilon} \approx 0$ for {linear} and branched structures. 
The energies at $k_B T = 1 \revision{\epsilon}$ and
$k_B T \approx 0$ differ by roughly $N/2$, indicating that this energy
difference can be attributed to thermal out-of-plane fluctuations of
vertices. 
Interestingly, the elastic energy of branched structures is 
found to be {\em lower} than
that of {linear} structures \revision{(Fig. \ref{fig:fig2}a, lower panel)}.  

To analyze this in more detail, we calculate separately the excess
elastic energy of caps (tube ends) and junctions {relative} to  
{a reference cylindrical tube section with the same radius and 
the same area (see Fig. \ref{fig:fig2}c)}: We
separate the structures into ``caps'', ``junctions'', and ``tubular''
sections as indicated in Fig.\ \ref{fig:fig2}b), extract an elastic
energy $e$ per tube length from the tubular sections, and evaluate the
excess energies of caps and junctions via $\Delta
E_{\text{cap,junction}}= E_{\text{cap,junction}}-l_{\text{ref}} \: e$,
where $l_{\text{ref}}=A_{\text{cap,junction}}/2 \pi R_T$ is the length
of {the} reference tube section. {For example, the} ideal values 
{for semispherical caps} are   
$e = \pi \kappa/R_{T}$ and $\Delta E_{\text{cap}}=3 \pi\kappa$,
{and this is independent of the cutoff value $r_c$ marking the
end of the ``cap'' region as long as $r_c > R_{T}$} . The procedure {thus} 
\revision{largely} removes the dependence of the
results on the specific dissection into junctions, caps, and tubular
regions.

{In practice,} the results are still somewhat sensitive to the choice 
of the cutoff {values} $r_c$ \revision{(Fig.\
\ref{fig:fig2}d, shaded areas)}. {Even taking these}
uncertainties into account, it is clear that the excess energy of caps 
is positive ($\Delta E_{\text{cap}}$ = 190-220\revision{$\epsilon$}
depending on the  applied force) and the excess energy of junctions
is negative ($\Delta E_{\text{junction}}\approx -280 \revision{\epsilon} $). 
The excess energy of caps is higher
than the theoretical estimate $E_{\text{cap}} = 3 \pi \kappa$, which
we attribute to some extra distortion in the vicinity of the vertex
where the \revision{force} $F_{\text{ext}}$ is applied. The negative
excess energy
of junctions \revision{reflects} the fact that the overall curvature
in the region of the junction is reduced. {Interestingly,} in 
branched structures, the energy gain at junctions more
than compensates the energy loss due to the formation of an additional
cap. As a result, the total elastic energy of branched structures is
lower than that of {linear structures}.


\begin{figure}	
		\centering
		\includegraphics[width=1.05\linewidth]
        {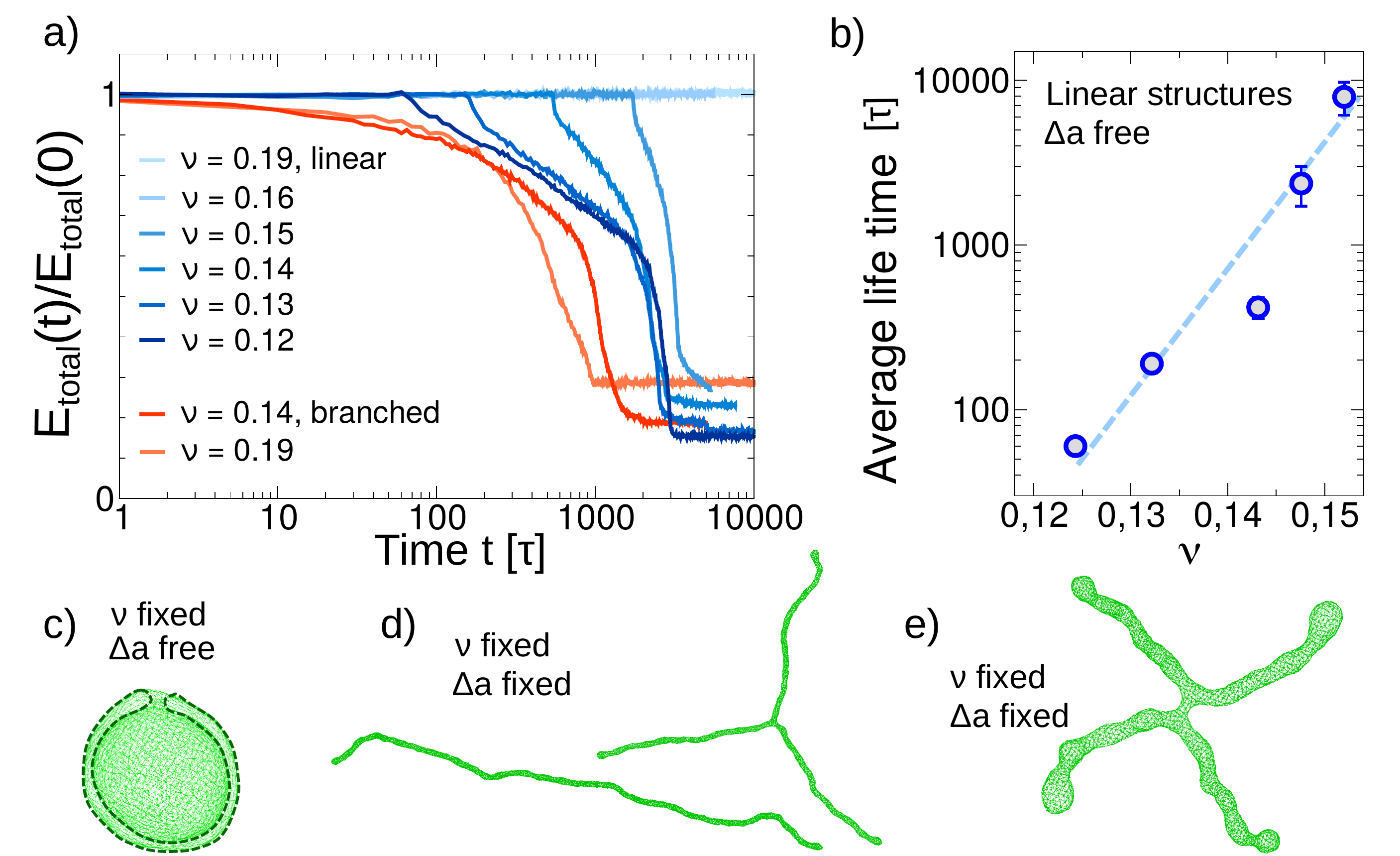}
	\caption{
    {Stability of force-free tubular structures.
    (a) Examples of time evolution of the total energy  after 
    releasing the force on force-stabilized 
    linear (blue) and branched  (red) structures at fixed $\nu$ 
    as indicated ($\Delta a$ is not constrained). 
    Both structures eventually transform into a stomatocyte. 
    For linear structures, the transformation process sets in after an
    activation time, which
    exceeds the maximum simulation time \revision{for $\nu > 0.15$}. 
    (b) Average life time of linear structures after releasing 
    the stabilizing force as a function of $\nu$ ($\Delta a$ is free). 
    Dashed line shows exponential behavior.
    (c) Example of a stable stomatocyte structure with $\nu=0.14$. 
    (d) Stable linear and branched structure if both $\nu$
    and $\Delta a$ are fixed ($\nu = 0.14$).
    (e) Structure obtained after releasing the force from a
    force-stabilized tetrahedral structure at fixed $\nu$ and 
    $\Delta a$ (at $\nu = 0.2$). The tetra-junction splits up 
    into two Y-junctions. 
    }
	\label{fig:fig3}
    }
\end{figure}


{\em Force-free (meta)stable structures.} We turn to the question
whether branched structures can be metastable in the absence of
forcing. Bahrami et al \cite{TriangulationBahrami2017} have recently
observed that {linear structures} remain metastable if the enclosed 
volume $V_0$ is fixed. Motivated by their findings, we study in Fig.\
\ref{fig:fig3}a) the time evolution of {linear} and branched 
structures after releasing a stabilizing force \revision{while 
keeping $V_0$ fixed.} 
In both cases, the structures eventually transform into 
{a structure with lower energy}, a stomatocyte
{(see Fig. \ref{fig:fig3}c)}. However, the transformation process 
is qualitatively different. {In branched structures}, it sets in 
immediately {via a disc-like widening at the junction 
(see SI movie {\tt branch\_fixNu.mp4})}. {Linear structures 
initially remain  (meta)stable for some activation time, 
indicating that the  shape change is an activated process. The
transformation} is then initiated by  the  nucleation of  a disc at 
one end {(see SI movie {\tt linear\_fixNu.mp4})}. 
{The activation time of linear structures increases roughly
exponentially   with $\nu$ (see Fig. \ref{fig:fig3}b) and 
eventually exceeds the total simulation time, consistent} with 
\cite{TriangulationBahrami2017}. 

Imposing a small reduced volume alone is thus not sufficient to
stabilize branched structures. However, constraining
the {average curvature} $\Delta a$ {in addition to $\nu$} does have a stabilizing effect. 
{SI Fig.\ 2a,b} shows that $\Delta a$ drops substantially during the
transformation from tubular/branched structures to stomatocyte. If
one constrains $\Delta a$ to its initial value, {i.e., the value of the force-stabilized structure}, this suppresses the transformation, and 
the tubular/branched structures with Y-junctions remain (meta)stable. 
{Examples are shown in Fig.\ \ref{fig:fig3}d).} 
Tetrahedral junctions, on the other hand, do not persist,  
but separate into two Y-junctions 
(see Fig.\ \ref{fig:fig3}d, SI Fig.\ 2e {and SI movie
{\tt f0\_tetrahedral\_nu20fixDa.mp4}}).
If only $\Delta a$ is kept fixed, linear
and branched structures are also stable, but may acquire
slightly pearled shapes, see SI Fig. 2d.

{\em Origin of energy penalty for junction defects}.
The question remains which of the two structures, branched or {linear}, 
has the lower energy.  Judging from our previous results on
excess cap and junction energies (Fig.\ \ref{fig:fig2}), one might
suspect that branching is energetically favorable. However, the
situation is more subtle. Adding a junction locally removes curvature
in the junction region, which has to be added elsewhere {to 
keep $\Delta a$ fixed}.  As a result, the tubular sections become 
thinner, and their curvature energy increases.  In SI, we present 
a theoretical estimate \revision{showing} that the resulting net 
energy difference 
is roughly given by  
$E_{_{\text{CV}},\text{branch}}- E_{_{\text{CV}},\text{linear}}
\sim \pi \kappa   + | \:  \varepsilon \cdot \Delta
E_{_{\text{CV}},\text{junction}}| {> 0}$, where $\varepsilon$ 
characterizes the reduction of curvature at the
junction. {At fixed $\Delta a$, the } total curvature energy of 
branched structures {should hence be} higher than that of 
{linear structures}.  A similar effect is expected 
\revision{for fixed $\nu$}:
A junction adds enclosed volume, which has to 
be removed elsewhere, 
leading again to a thinning of tubes.  

The net effect of constraints on the curvature energy as obtained from
simulations is summarized in table \ref{tab:constraints} for the
example of force-stabilized structures at $F_{\text{ext}}=70 \revision{\epsilon/l_b}$.
Here, we have used the values of $\nu$ and/or $\Delta a$ obtained 
for unconstrained force-stabilized linear structures (parameter set C1)
and branched structures (parameter set C2) as input parameters in constrained force-stabilized simulations of linear and branched 
structures. \revision{The curvature energies obtained with the set C1 are generally higher than those obtained with C2, because 
$\Delta a$ is higher and/or $\nu$ is lower. Comparing linear
and branched structures for the same parameter set, the} 
results confirm the expectations of the discussion above. Only in the absence of any constraints is the curvature energy of {branched structures} lower than that of {linear structures}.  
In all other cases (constraints on $\nu$, on $\Delta a$, on both), 
the curvature energy of {branched structures} is higher.

\begin{table}
	\centering
	{
	\begin{tabular}[t]{l|l|c|c|c|c|c|c}
		\hline
        Fixed
        & Structure
        & \multicolumn{2}{c}{ $E_{CV}\revision{/\epsilon}$ }& \multicolumn{2}{c}{$\nu$} & \multicolumn{2}{c}{$\Delta a$} \\
        \hline
        -- & {Linear} &
        \multicolumn{2}{c}{$ 9333 \pm \gj{26}$} &  
        \multicolumn{2}{c}{$0.186  \pm \gj{0.001} $} & \multicolumn{2}{c}{$3.90   \pm \gj{0.01}$} \\
        & Branched &
        \multicolumn{2}{c}{$9082 \pm \gj{34} $} &  
        \multicolumn{2}{c}{$0.193   \pm \gj{0.001} $} & \multicolumn{2}{c}{ $3.81  \pm \gj{0.01}$} \\
        \hline
        &&\multicolumn{3}{c|}{C1 ({Linear})} &\multicolumn{3}{c}{C2 (Branched)}\\
        &&$E_{CV}$&$\nu$&$\Delta a$ & $E_{CV}$&$\nu$&$\Delta a$\\
        \hline
        $\Delta a$ & {Linear} &
            \gj    {$9333 \pm 3$} &
       \gj{0.186} & \gj{3.90}  &
        $8975 \pm 3$ &
        0.191 & 3.81 \\ 
        & Branched &
        
                \gj{$9470 \pm 3$} &
       \gj{0.187}  & \gj{3.90}&
        
        $9117 \pm 3$ &
        0.192 & 3.81\\
		\hline
        $\nu$ & {Linear} &
        \gj{$9355 \pm 3$} &
        \gj{0.186} & \gj{3.91} &
        $8882 \pm 3$ &
        0.192 & 3.79\\
        & Branch &
        
        \gj{9538} $\pm 4$ &
        \gj{0.186} & \gj{3.92} &
       
        $9079 \pm 4$ &
        0.193 &  3.80\\
        \hline
        $\nu, \Delta a$ & {Linear}&
        $9342 \pm 3$ &
       0.186  & 3.90 &
        $8951 \pm 3$ &
        0.192 & 3.81 \\
        & Branch &
        $9480 \pm 3$ &
       0.186 & 3.90  &
        $9126 \pm 3$ &
       0.193 & 3.81 \\
		\hline 
	\end{tabular}
	}
	\caption{Curvature energies $E_{CV}$, reduced volumes $\nu$, and {average	curvatures $\Delta a$} for force-stabilized pure {linear} and branched structures at $F_{\text{ext}}=70 \revision{\epsilon/l_b}$. {Results are shown for two sets of constraints C1 and C2
	on $\nu$, $\Delta a$, or both as indicated, which correspond to 
	the values obtained for unconstrained force-stabilized linear 
	and branched structures, respectively.}} 
	\label{tab:constraints}
\end{table}

{\em Conclusions.} To summarize, we have investigated the energetics
and stability of an essential component of tubular membrane networks,
the junctions, from the point of view of the Canham-Helfrich elastic
theory of membranes.  We consider membrane structures with closed
sphere topology and allow for constraints on the enclosed volume $\nu$
and the {average curvature} $\Delta a$, without however imposing
specific local curvatures.  Within this simple model, we find that 
Y-junctions with angles $120^o$ can be stabilized by mechanical forces
and remain metastable after releasing the forces. Other types of
junctions and other angles are unstable. Furthermore, we find that
Y-junctions locally have a negative excess curvature energy. For fixed tube
diameter, branching is energetically favorable, even if one accounts
for the positive energy of the additional cap. At fixed $\Delta a$,
however, adding a branch enforces a thinning of the tubes, such that
the overall curvature energy balance disfavors branching. 

This subtle energy balance should lead to an increase of the lifetime
of metastable branches, as their elimination is only favorable if
the entire tube network rearranges. In addition, dynamical simulations
suggest that the creation and annihilation of branches is accompanied
by a free energy barrier: Pulling a branch out of a tube requires
slightly higher forces than needed to stabilize it (see SI, Fig.\ 2a)
and if one annihilates a branch by pulling on the
other tubes, the curvature energy passes through a maximum 
(SI, Fig.\ 2b)

Our results thus indicate that simple properties of elastic membranes
might be responsible for the abundance of tube network structures in
cells. These structures are already metastable and long-lived if one
imposes a few generic constraints, such as a fixed {surface} area 
difference between inner and outer membrane leaflet and {possibly} 
impermeability (fixed enclosed volume; {not strictly necessary}). 
Hence they can be  stabilized and manipulated with little extra effort.

\revision{We have studied a very idealized model of bare membranes. 
However, given the generic character of our main conclusions, we expect 
them to still hold in other membrane models, e.g., ADE models with 
more realistic (lower) area  difference elasticities $k_{\Delta a}$, 
or membrane structures with average curvature imposed by freely 
moving curvature-inducing proteins \cite{Tsafrir2003,Shibata2009,Campelo2008,Frost2009,Voeltz2006,Hu2008}, 
where one has to account for their entropy of mixing. 
This will be an interesting subject for future studies. }


We thank Enrico Schleiff for motivating this project, Hiroshi Noguchi 
from the ISSP at the University of Tokyo, Japan, for helpful discussions 
and for sharing the source code of his dynamically-triangulated
membrane model \cite{TriangulationNoguchi2005}, which was extended by
us  to include area difference constraints. The simulations
were carried out on the supercomputer Mogon at Johannes Gutenberg
University Mainz.  This work was funded by the state of
Rhineland-Palatinate, Germany, within the Dynamem consortium, and in
part by the Deutsche Forschungsgemeinschaft (DFG) via Grant 233630050
(SFB TRR 146). 



\begin{thebibliography}{69}%
\makeatletter
\providecommand \@ifxundefined [1]{%
 \@ifx{#1\undefined}
}%
\providecommand \@ifnum [1]{%
 \ifnum #1\expandafter \@firstoftwo
 \else \expandafter \@secondoftwo
 \fi
}%
\providecommand \@ifx [1]{%
 \ifx #1\expandafter \@firstoftwo
 \else \expandafter \@secondoftwo
 \fi
}%
\providecommand \natexlab [1]{#1}%
\providecommand \enquote  [1]{``#1''}%
\providecommand \bibnamefont  [1]{#1}%
\providecommand \bibfnamefont [1]{#1}%
\providecommand \citenamefont [1]{#1}%
\providecommand \href@noop [0]{\@secondoftwo}%
\providecommand \href [0]{\begingroup \@sanitize@url \@href}%
\providecommand \@href[1]{\@@startlink{#1}\@@href}%
\providecommand \@@href[1]{\endgroup#1\@@endlink}%
\providecommand \@sanitize@url [0]{\catcode `\\12\catcode `\$12\catcode
  `\&12\catcode `\#12\catcode `\^12\catcode `\_12\catcode `\%12\relax}%
\providecommand \@@startlink[1]{}%
\providecommand \@@endlink[0]{}%
\providecommand \url  [0]{\begingroup\@sanitize@url \@url }%
\providecommand \@url [1]{\endgroup\@href {#1}{\urlprefix }}%
\providecommand \urlprefix  [0]{URL }%
\providecommand \Eprint [0]{\href }%
\providecommand \doibase [0]{http://dx.doi.org/}%
\providecommand \selectlanguage [0]{\@gobble}%
\providecommand \bibinfo  [0]{\@secondoftwo}%
\providecommand \bibfield  [0]{\@secondoftwo}%
\providecommand \translation [1]{[#1]}%
\providecommand \BibitemOpen [0]{}%
\providecommand \bibitemStop [0]{}%
\providecommand \bibitemNoStop [0]{.\EOS\space}%
\providecommand \EOS [0]{\spacefactor3000\relax}%
\providecommand \BibitemShut  [1]{\csname bibitem#1\endcsname}%
\let\auto@bib@innerbib\@empty
\bibitem [{\citenamefont {Mollenhauer}\ and\ \citenamefont
  {Morre}(1998)}]{Mollenhauer1998}%
  \BibitemOpen
  \bibfield  {author} {\bibinfo {author} {\bibfnamefont {H.}~\bibnamefont
  {Mollenhauer}}\ and\ \bibinfo {author} {\bibfnamefont {D.}~\bibnamefont
  {Morre}},\ }\href@noop {} {\bibfield  {journal} {\bibinfo  {journal}
  {Histochemistry and Cell Biology}\ }\textbf {\bibinfo {volume} {109}},\
  \bibinfo {pages} {533} (\bibinfo {year} {1998})}\BibitemShut {NoStop}%
\bibitem [{\citenamefont {De~Matteis}\ and\ \citenamefont
  {Luini}(2008)}]{MatteisGolgi2008}%
  \BibitemOpen
  \bibfield  {author} {\bibinfo {author} {\bibfnamefont {M.}~\bibnamefont
  {De~Matteis}}\ and\ \bibinfo {author} {\bibfnamefont {A.}~\bibnamefont
  {Luini}},\ }\href {\doibase 10.1038/nrm2378} {\bibfield  {journal} {\bibinfo
  {journal} {Nature reviews. Molecular cell biology}\ }\textbf {\bibinfo
  {volume} {9}},\ \bibinfo {pages} {273} (\bibinfo {year} {2008})}\BibitemShut
  {NoStop}%
\bibitem [{\citenamefont {Westrate}\ \emph {et~al.}(2015)\citenamefont
  {Westrate}, \citenamefont {Lee}, \citenamefont {Prinz},\ and\ \citenamefont
  {Voeltz}}]{WestrateER2015}%
  \BibitemOpen
  \bibfield  {author} {\bibinfo {author} {\bibfnamefont {L.}~\bibnamefont
  {Westrate}}, \bibinfo {author} {\bibfnamefont {J.}~\bibnamefont {Lee}},
  \bibinfo {author} {\bibfnamefont {W.}~\bibnamefont {Prinz}}, \ and\ \bibinfo
  {author} {\bibfnamefont {G.}~\bibnamefont {Voeltz}},\ }\href {\doibase
  10.1146/annurev-biochem-072711-163501} {\bibfield  {journal} {\bibinfo
  {journal} {Annual Review of Biochemistry}\ }\textbf {\bibinfo {volume}
  {84}},\ \bibinfo {pages} {791} (\bibinfo {year} {2015})},\ \bibinfo {note}
  {pMID: 25580528},\ \Eprint
  {http://arxiv.org/abs/https://doi.org/10.1146/annurev-biochem-072711-163501}
  {https://doi.org/10.1146/annurev-biochem-072711-163501} \BibitemShut
  {NoStop}%
\bibitem [{\citenamefont {Lü}\ \emph {et~al.}(2020)\citenamefont {Lü},
  \citenamefont {Niu},\ and\ \citenamefont {Hu}}]{Lue2020}%
  \BibitemOpen
  \bibfield  {author} {\bibinfo {author} {\bibfnamefont {L.}~\bibnamefont
  {Lü}}, \bibinfo {author} {\bibfnamefont {L.}~\bibnamefont {Niu}}, \ and\
  \bibinfo {author} {\bibfnamefont {J.}~\bibnamefont {Hu}},\ }\href {\doibase
  10.1007/s41048-020-00113-y} {\bibfield  {journal} {\bibinfo  {journal}
  {Biophysics Reports}\ }\textbf {\bibinfo {volume} {6}} (\bibinfo {year}
  {2020}),\ 10.1007/s41048-020-00113-y}\BibitemShut {NoStop}%
\bibitem [{\citenamefont {Powers}\ \emph {et~al.}(2017)\citenamefont {Powers},
  \citenamefont {Wang}, \citenamefont {Liu},\ and\ \citenamefont
  {Rapoport}}]{Powers2017}%
  \BibitemOpen
  \bibfield  {author} {\bibinfo {author} {\bibfnamefont {R.~E.}\ \bibnamefont
  {Powers}}, \bibinfo {author} {\bibfnamefont {S.}~\bibnamefont {Wang}},
  \bibinfo {author} {\bibfnamefont {T.~Y.}\ \bibnamefont {Liu}}, \ and\
  \bibinfo {author} {\bibfnamefont {T.~A.}\ \bibnamefont {Rapoport}},\
  }\href@noop {} {\bibfield  {journal} {\bibinfo  {journal} {Nature}\ }\textbf
  {\bibinfo {volume} {543}},\ \bibinfo {pages} {257} (\bibinfo {year}
  {2017})}\BibitemShut {NoStop}%
\bibitem [{\citenamefont {Saraste}\ and\ \citenamefont
  {Prydz}(2019)}]{Saraste2019}%
  \BibitemOpen
  \bibfield  {author} {\bibinfo {author} {\bibfnamefont {J.}~\bibnamefont
  {Saraste}}\ and\ \bibinfo {author} {\bibfnamefont {K.}~\bibnamefont
  {Prydz}},\ }\href {\doibase 10.3389/fcell.2019.00171} {\bibfield  {journal}
  {\bibinfo  {journal} {Frontiers in Cell and Developmental Biology}\ }\textbf
  {\bibinfo {volume} {7}},\ \bibinfo {pages} {171} (\bibinfo {year}
  {2019})}\BibitemShut {NoStop}%
\bibitem [{\citenamefont {Sowinski}\ \emph {et~al.}(2008)\citenamefont
  {Sowinski}, \citenamefont {Jolly}, \citenamefont {Berninghausen},
  \citenamefont {Purbhoo}, \citenamefont {Chauveau}, \citenamefont
  {K{\"o}hler}, \citenamefont {Oddos}, \citenamefont {Eissmann}, \citenamefont
  {Brodsky}, \citenamefont {Hopkins}, \citenamefont {{\"O}nfelt}, \citenamefont
  {Sattentau},\ and\ \citenamefont {Davis}}]{Sowinski2008}%
  \BibitemOpen
  \bibfield  {author} {\bibinfo {author} {\bibfnamefont {S.}~\bibnamefont
  {Sowinski}}, \bibinfo {author} {\bibfnamefont {C.}~\bibnamefont {Jolly}},
  \bibinfo {author} {\bibfnamefont {O.}~\bibnamefont {Berninghausen}}, \bibinfo
  {author} {\bibfnamefont {M.}~\bibnamefont {Purbhoo}}, \bibinfo {author}
  {\bibfnamefont {A.}~\bibnamefont {Chauveau}}, \bibinfo {author}
  {\bibfnamefont {K.}~\bibnamefont {K{\"o}hler}}, \bibinfo {author}
  {\bibfnamefont {S.}~\bibnamefont {Oddos}}, \bibinfo {author} {\bibfnamefont
  {P.}~\bibnamefont {Eissmann}}, \bibinfo {author} {\bibfnamefont
  {F.}~\bibnamefont {Brodsky}}, \bibinfo {author} {\bibfnamefont
  {C.}~\bibnamefont {Hopkins}}, \bibinfo {author} {\bibfnamefont
  {B.}~\bibnamefont {{\"O}nfelt}}, \bibinfo {author} {\bibfnamefont
  {Q.}~\bibnamefont {Sattentau}}, \ and\ \bibinfo {author} {\bibfnamefont
  {D.}~\bibnamefont {Davis}},\ }\href {\doibase 10.1038/ncb1682} {\bibfield
  {journal} {\bibinfo  {journal} {Nature Cell Biology}\ }\textbf {\bibinfo
  {volume} {10}},\ \bibinfo {pages} {211} (\bibinfo {year} {2008})}\BibitemShut
  {NoStop}%
\bibitem [{\citenamefont {Roux}(2013)}]{Roux2013}%
  \BibitemOpen
  \bibfield  {author} {\bibinfo {author} {\bibfnamefont {A.}~\bibnamefont
  {Roux}},\ }\href {\doibase 10.1039/C3SM50514F} {\bibfield  {journal}
  {\bibinfo  {journal} {Soft Matter}\ }\textbf {\bibinfo {volume} {9}},\
  \bibinfo {pages} {6726} (\bibinfo {year} {2013})}\BibitemShut {NoStop}%
\bibitem [{\citenamefont {Miyata}\ and\ \citenamefont
  {Hotani}(1992)}]{Miyata1992}%
  \BibitemOpen
  \bibfield  {author} {\bibinfo {author} {\bibfnamefont {H.}~\bibnamefont
  {Miyata}}\ and\ \bibinfo {author} {\bibfnamefont {H.}~\bibnamefont
  {Hotani}},\ }\href {\doibase 10.1073/pnas.89.23.11547} {\bibfield  {journal}
  {\bibinfo  {journal} {Proceedings of the National Academy of Sciences}\
  }\textbf {\bibinfo {volume} {89}},\ \bibinfo {pages} {11547} (\bibinfo {year}
  {1992})},\ \Eprint
  {http://arxiv.org/abs/https://www.pnas.org/content/89/23/11547.full.pdf}
  {https://www.pnas.org/content/89/23/11547.full.pdf} \BibitemShut {NoStop}%
\bibitem [{\citenamefont {Miyata}\ \emph {et~al.}(1999)\citenamefont {Miyata},
  \citenamefont {Nishiyama}, \citenamefont {Akashi},\ and\ \citenamefont
  {Kinosita}}]{Miyata1999}%
  \BibitemOpen
  \bibfield  {author} {\bibinfo {author} {\bibfnamefont {H.}~\bibnamefont
  {Miyata}}, \bibinfo {author} {\bibfnamefont {S.}~\bibnamefont {Nishiyama}},
  \bibinfo {author} {\bibfnamefont {K.-i.}\ \bibnamefont {Akashi}}, \ and\
  \bibinfo {author} {\bibfnamefont {K.}~\bibnamefont {Kinosita}},\ }\href
  {\doibase 10.1073/pnas.96.5.2048} {\bibfield  {journal} {\bibinfo  {journal}
  {Proceedings of the National Academy of Sciences}\ }\textbf {\bibinfo
  {volume} {96}},\ \bibinfo {pages} {2048} (\bibinfo {year} {1999})},\ \Eprint
  {http://arxiv.org/abs/https://www.pnas.org/content/96/5/2048.full.pdf}
  {https://www.pnas.org/content/96/5/2048.full.pdf} \BibitemShut {NoStop}%
\bibitem [{\citenamefont {Der\'enyi}\ \emph {et~al.}(2002)\citenamefont
  {Der\'enyi}, \citenamefont {J\"ulicher},\ and\ \citenamefont
  {Prost}}]{DerenyiProst2002}%
  \BibitemOpen
  \bibfield  {author} {\bibinfo {author} {\bibfnamefont {I.}~\bibnamefont
  {Der\'enyi}}, \bibinfo {author} {\bibfnamefont {F.}~\bibnamefont
  {J\"ulicher}}, \ and\ \bibinfo {author} {\bibfnamefont {J.}~\bibnamefont
  {Prost}},\ }\href {\doibase 10.1103/PhysRevLett.88.238101} {\bibfield
  {journal} {\bibinfo  {journal} {Phys. Rev. Lett.}\ }\textbf {\bibinfo
  {volume} {88}},\ \bibinfo {pages} {238101} (\bibinfo {year}
  {2002})}\BibitemShut {NoStop}%
\bibitem [{\citenamefont {Koster}\ \emph {et~al.}(2003)\citenamefont {Koster},
  \citenamefont {VanDuijn}, \citenamefont {Hofs},\ and\ \citenamefont
  {Dogterom}}]{Koster2003}%
  \BibitemOpen
  \bibfield  {author} {\bibinfo {author} {\bibfnamefont {G.}~\bibnamefont
  {Koster}}, \bibinfo {author} {\bibfnamefont {M.}~\bibnamefont {VanDuijn}},
  \bibinfo {author} {\bibfnamefont {B.}~\bibnamefont {Hofs}}, \ and\ \bibinfo
  {author} {\bibfnamefont {M.}~\bibnamefont {Dogterom}},\ }\href {\doibase
  10.1073/pnas.2531786100} {\bibfield  {journal} {\bibinfo  {journal}
  {Proceedings of the National Academy of Sciences}\ }\textbf {\bibinfo
  {volume} {100}},\ \bibinfo {pages} {15583} (\bibinfo {year} {2003})},\
  \Eprint
  {http://arxiv.org/abs/https://www.pnas.org/content/100/26/15583.full.pdf}
  {https://www.pnas.org/content/100/26/15583.full.pdf} \BibitemShut {NoStop}%
\bibitem [{\citenamefont {Leduc}\ \emph {et~al.}(2004)\citenamefont {Leduc},
  \citenamefont {Camp{\`a}s}, \citenamefont {Zeldovich}, \citenamefont {Roux},
  \citenamefont {Jolimaitre}, \citenamefont {Bourel-Bonnet}, \citenamefont
  {Goud}, \citenamefont {Joanny}, \citenamefont {Bassereau},\ and\
  \citenamefont {Prost}}]{Leduc2004}%
  \BibitemOpen
  \bibfield  {author} {\bibinfo {author} {\bibfnamefont {C.}~\bibnamefont
  {Leduc}}, \bibinfo {author} {\bibfnamefont {O.}~\bibnamefont {Camp{\`a}s}},
  \bibinfo {author} {\bibfnamefont {K.~B.}\ \bibnamefont {Zeldovich}}, \bibinfo
  {author} {\bibfnamefont {A.}~\bibnamefont {Roux}}, \bibinfo {author}
  {\bibfnamefont {P.}~\bibnamefont {Jolimaitre}}, \bibinfo {author}
  {\bibfnamefont {L.}~\bibnamefont {Bourel-Bonnet}}, \bibinfo {author}
  {\bibfnamefont {B.}~\bibnamefont {Goud}}, \bibinfo {author} {\bibfnamefont
  {J.-F.}\ \bibnamefont {Joanny}}, \bibinfo {author} {\bibfnamefont
  {P.}~\bibnamefont {Bassereau}}, \ and\ \bibinfo {author} {\bibfnamefont
  {J.}~\bibnamefont {Prost}},\ }\href {\doibase 10.1073/pnas.0406598101}
  {\bibfield  {journal} {\bibinfo  {journal} {Proceedings of the National
  Academy of Sciences}\ }\textbf {\bibinfo {volume} {101}},\ \bibinfo {pages}
  {17096} (\bibinfo {year} {2004})},\ \Eprint
  {http://arxiv.org/abs/https://www.pnas.org/content/101/49/17096.full.pdf}
  {https://www.pnas.org/content/101/49/17096.full.pdf} \BibitemShut {NoStop}%
\bibitem [{\citenamefont {Campàs}\ \emph {et~al.}(2008)\citenamefont
  {Campàs}, \citenamefont {Leduc}, \citenamefont {Bassereau}, \citenamefont
  {Casademunt}, \citenamefont {Joanny},\ and\ \citenamefont
  {Prost}}]{Campas2008}%
  \BibitemOpen
  \bibfield  {author} {\bibinfo {author} {\bibfnamefont {O.}~\bibnamefont
  {Campàs}}, \bibinfo {author} {\bibfnamefont {C.}~\bibnamefont {Leduc}},
  \bibinfo {author} {\bibfnamefont {P.}~\bibnamefont {Bassereau}}, \bibinfo
  {author} {\bibfnamefont {J.}~\bibnamefont {Casademunt}}, \bibinfo {author}
  {\bibfnamefont {J.-F.}\ \bibnamefont {Joanny}}, \ and\ \bibinfo {author}
  {\bibfnamefont {J.}~\bibnamefont {Prost}},\ }\href {\doibase
  https://doi.org/10.1529/biophysj.107.118554} {\bibfield  {journal} {\bibinfo
  {journal} {Biophysical Journal}\ }\textbf {\bibinfo {volume} {94}},\ \bibinfo
  {pages} {5009 } (\bibinfo {year} {2008})}\BibitemShut {NoStop}%
\bibitem [{\citenamefont {Nambiar}\ \emph {et~al.}(2009)\citenamefont
  {Nambiar}, \citenamefont {E.},\ and\ \citenamefont {Tyska}}]{Nambiar2009}%
  \BibitemOpen
  \bibfield  {author} {\bibinfo {author} {\bibfnamefont {R.}~\bibnamefont
  {Nambiar}}, \bibinfo {author} {\bibfnamefont {M.~R.}\ \bibnamefont {E.}}, \
  and\ \bibinfo {author} {\bibfnamefont {M.~J.}\ \bibnamefont {Tyska}},\ }\href
  {\doibase https://doi.org/10.1073/pnas.0901641106} {\bibfield  {journal}
  {\bibinfo  {journal} {PNAS}\ }\textbf {\bibinfo {volume} {106}},\ \bibinfo
  {pages} {11972} (\bibinfo {year} {2009})}\BibitemShut {NoStop}%
\bibitem [{\citenamefont {Du}\ \emph {et~al.}(2016)\citenamefont {Du},
  \citenamefont {Su}, \citenamefont {Chen}, \citenamefont {Zhu}, \citenamefont
  {Jiang}, \citenamefont {Rong}, \citenamefont {Zhang}, \citenamefont {Zhang},
  \citenamefont {Ren}, \citenamefont {Zhang}, \citenamefont {Wang},
  \citenamefont {Gao}, \citenamefont {Wang}, \citenamefont {Sun}, \citenamefont
  {Sun},\ and\ \citenamefont {Yu}}]{Du2016}%
  \BibitemOpen
  \bibfield  {author} {\bibinfo {author} {\bibfnamefont {W.}~\bibnamefont
  {Du}}, \bibinfo {author} {\bibfnamefont {Q.}~\bibnamefont {Su}}, \bibinfo
  {author} {\bibfnamefont {Y.}~\bibnamefont {Chen}}, \bibinfo {author}
  {\bibfnamefont {Y.}~\bibnamefont {Zhu}}, \bibinfo {author} {\bibfnamefont
  {D.}~\bibnamefont {Jiang}}, \bibinfo {author} {\bibfnamefont
  {Y.}~\bibnamefont {Rong}}, \bibinfo {author} {\bibfnamefont {S.}~\bibnamefont
  {Zhang}}, \bibinfo {author} {\bibfnamefont {Y.}~\bibnamefont {Zhang}},
  \bibinfo {author} {\bibfnamefont {H.}~\bibnamefont {Ren}}, \bibinfo {author}
  {\bibfnamefont {C.}~\bibnamefont {Zhang}}, \bibinfo {author} {\bibfnamefont
  {X.}~\bibnamefont {Wang}}, \bibinfo {author} {\bibfnamefont {N.}~\bibnamefont
  {Gao}}, \bibinfo {author} {\bibfnamefont {Y.}~\bibnamefont {Wang}}, \bibinfo
  {author} {\bibfnamefont {L.}~\bibnamefont {Sun}}, \bibinfo {author}
  {\bibfnamefont {Y.}~\bibnamefont {Sun}}, \ and\ \bibinfo {author}
  {\bibfnamefont {L.}~\bibnamefont {Yu}},\ }\href {\doibase
  https://doi.org/10.1016/j.devcel.2016.04.014} {\bibfield  {journal} {\bibinfo
   {journal} {Developmental Cell}\ }\textbf {\bibinfo {volume} {37}},\ \bibinfo
  {pages} {326 } (\bibinfo {year} {2016})}\BibitemShut {NoStop}%
\bibitem [{\citenamefont {Footer}\ \emph {et~al.}(2007)\citenamefont {Footer},
  \citenamefont {Kerssemakers}, \citenamefont {Theriot},\ and\ \citenamefont
  {Dogterom}}]{Footer2007}%
  \BibitemOpen
  \bibfield  {author} {\bibinfo {author} {\bibfnamefont {M.~J.}\ \bibnamefont
  {Footer}}, \bibinfo {author} {\bibfnamefont {J.~W.~J.}\ \bibnamefont
  {Kerssemakers}}, \bibinfo {author} {\bibfnamefont {J.~A.}\ \bibnamefont
  {Theriot}}, \ and\ \bibinfo {author} {\bibfnamefont {M.}~\bibnamefont
  {Dogterom}},\ }\href {\doibase 10.1073/pnas.0607052104} {\bibfield  {journal}
  {\bibinfo  {journal} {Proceedings of the National Academy of Sciences}\
  }\textbf {\bibinfo {volume} {104}},\ \bibinfo {pages} {2181} (\bibinfo {year}
  {2007})},\ \Eprint
  {http://arxiv.org/abs/https://www.pnas.org/content/104/7/2181.full.pdf}
  {https://www.pnas.org/content/104/7/2181.full.pdf} \BibitemShut {NoStop}%
\bibitem [{\citenamefont {Roux}\ \emph {et~al.}(2010)\citenamefont {Roux},
  \citenamefont {Koster}, \citenamefont {Lenz}, \citenamefont {Sorre},
  \citenamefont {Manneville}, \citenamefont {Nassoy},\ and\ \citenamefont
  {Bassereau}}]{Roux2010}%
  \BibitemOpen
  \bibfield  {author} {\bibinfo {author} {\bibfnamefont {A.}~\bibnamefont
  {Roux}}, \bibinfo {author} {\bibfnamefont {G.}~\bibnamefont {Koster}},
  \bibinfo {author} {\bibfnamefont {M.}~\bibnamefont {Lenz}}, \bibinfo {author}
  {\bibfnamefont {B.}~\bibnamefont {Sorre}}, \bibinfo {author} {\bibfnamefont
  {J.-B.}\ \bibnamefont {Manneville}}, \bibinfo {author} {\bibfnamefont
  {P.}~\bibnamefont {Nassoy}}, \ and\ \bibinfo {author} {\bibfnamefont
  {P.}~\bibnamefont {Bassereau}},\ }\href {\doibase 10.1073/pnas.0913734107}
  {\bibfield  {journal} {\bibinfo  {journal} {Proceedings of the National
  Academy of Sciences}\ }\textbf {\bibinfo {volume} {107}},\ \bibinfo {pages}
  {4141} (\bibinfo {year} {2010})},\ \Eprint
  {http://arxiv.org/abs/https://www.pnas.org/content/107/9/4141.full.pdf}
  {https://www.pnas.org/content/107/9/4141.full.pdf} \BibitemShut {NoStop}%
\bibitem [{\citenamefont {Tsafrir}\ \emph {et~al.}(2003)\citenamefont
  {Tsafrir}, \citenamefont {Caspi}, \citenamefont {Guedeau-Boudeville},
  \citenamefont {Arzi},\ and\ \citenamefont {Stavans}}]{Tsafrir2003}%
  \BibitemOpen
  \bibfield  {author} {\bibinfo {author} {\bibfnamefont {I.}~\bibnamefont
  {Tsafrir}}, \bibinfo {author} {\bibfnamefont {Y.}~\bibnamefont {Caspi}},
  \bibinfo {author} {\bibfnamefont {M.-A.}\ \bibnamefont {Guedeau-Boudeville}},
  \bibinfo {author} {\bibfnamefont {T.}~\bibnamefont {Arzi}}, \ and\ \bibinfo
  {author} {\bibfnamefont {J.}~\bibnamefont {Stavans}},\ }\href {\doibase
  10.1103/PhysRevLett.91.138102} {\bibfield  {journal} {\bibinfo  {journal}
  {Phys. Rev. Lett.}\ }\textbf {\bibinfo {volume} {91}},\ \bibinfo {pages}
  {138102} (\bibinfo {year} {2003})}\BibitemShut {NoStop}%
\bibitem [{\citenamefont {Shibata}\ \emph {et~al.}(2009)\citenamefont
  {Shibata}, \citenamefont {Hu}, \citenamefont {Kozlov},\ and\ \citenamefont
  {Rapoport}}]{Shibata2009}%
  \BibitemOpen
  \bibfield  {author} {\bibinfo {author} {\bibfnamefont {Y.}~\bibnamefont
  {Shibata}}, \bibinfo {author} {\bibfnamefont {J.}~\bibnamefont {Hu}},
  \bibinfo {author} {\bibfnamefont {M.~M.}\ \bibnamefont {Kozlov}}, \ and\
  \bibinfo {author} {\bibfnamefont {T.~A.}\ \bibnamefont {Rapoport}},\ }\href
  {\doibase 10.1146/annurev.cellbio.042308.113324} {\bibfield  {journal}
  {\bibinfo  {journal} {Annual Review of Cell and Developmental Biology}\
  }\textbf {\bibinfo {volume} {25}},\ \bibinfo {pages} {329} (\bibinfo {year}
  {2009})}\BibitemShut {NoStop}%
\bibitem [{\citenamefont {Campelo}\ \emph {et~al.}(2008)\citenamefont
  {Campelo}, \citenamefont {McMahon},\ and\ \citenamefont
  {Kozlov}}]{Campelo2008}%
  \BibitemOpen
  \bibfield  {author} {\bibinfo {author} {\bibfnamefont {F.}~\bibnamefont
  {Campelo}}, \bibinfo {author} {\bibfnamefont {H.~T.}\ \bibnamefont
  {McMahon}}, \ and\ \bibinfo {author} {\bibfnamefont {M.~M.}\ \bibnamefont
  {Kozlov}},\ }\href {\doibase https://doi.org/10.1529/biophysj.108.133173}
  {\bibfield  {journal} {\bibinfo  {journal} {Biophysical Journal}\ }\textbf
  {\bibinfo {volume} {95}},\ \bibinfo {pages} {2325 } (\bibinfo {year}
  {2008})}\BibitemShut {NoStop}%
\bibitem [{\citenamefont {Frost}\ \emph {et~al.}(2009)\citenamefont {Frost},
  \citenamefont {Unger},\ and\ \citenamefont {{De Camilli}}}]{Frost2009}%
  \BibitemOpen
  \bibfield  {author} {\bibinfo {author} {\bibfnamefont {A.}~\bibnamefont
  {Frost}}, \bibinfo {author} {\bibfnamefont {V.~M.}\ \bibnamefont {Unger}}, \
  and\ \bibinfo {author} {\bibfnamefont {P.}~\bibnamefont {{De Camilli}}},\
  }\href {\doibase https://doi.org/10.1016/j.cell.2009.04.010} {\bibfield
  {journal} {\bibinfo  {journal} {Cell}\ }\textbf {\bibinfo {volume} {137}},\
  \bibinfo {pages} {191 } (\bibinfo {year} {2009})}\BibitemShut {NoStop}%
\bibitem [{\citenamefont {Voeltz}\ \emph {et~al.}(2006)\citenamefont {Voeltz},
  \citenamefont {Prinz}, \citenamefont {Shibata}, \citenamefont {Rist},\ and\
  \citenamefont {Rapoport}}]{Voeltz2006}%
  \BibitemOpen
  \bibfield  {author} {\bibinfo {author} {\bibfnamefont {G.~K.}\ \bibnamefont
  {Voeltz}}, \bibinfo {author} {\bibfnamefont {W.~A.}\ \bibnamefont {Prinz}},
  \bibinfo {author} {\bibfnamefont {Y.}~\bibnamefont {Shibata}}, \bibinfo
  {author} {\bibfnamefont {J.~M.}\ \bibnamefont {Rist}}, \ and\ \bibinfo
  {author} {\bibfnamefont {T.~A.}\ \bibnamefont {Rapoport}},\ }\href {\doibase
  https://doi.org/10.1016/j.cell.2005.11.047} {\bibfield  {journal} {\bibinfo
  {journal} {Cell}\ }\textbf {\bibinfo {volume} {124}},\ \bibinfo {pages} {573
  } (\bibinfo {year} {2006})}\BibitemShut {NoStop}%
\bibitem [{\citenamefont {Hu}\ \emph {et~al.}(2008)\citenamefont {Hu},
  \citenamefont {Shibata}, \citenamefont {Voss}, \citenamefont {Shemesh},
  \citenamefont {Li}, \citenamefont {Coughlin}, \citenamefont {Kozlov},
  \citenamefont {Rapoport},\ and\ \citenamefont {Prinz}}]{Hu2008}%
  \BibitemOpen
  \bibfield  {author} {\bibinfo {author} {\bibfnamefont {J.}~\bibnamefont
  {Hu}}, \bibinfo {author} {\bibfnamefont {Y.}~\bibnamefont {Shibata}},
  \bibinfo {author} {\bibfnamefont {C.}~\bibnamefont {Voss}}, \bibinfo {author}
  {\bibfnamefont {T.}~\bibnamefont {Shemesh}}, \bibinfo {author} {\bibfnamefont
  {Z.}~\bibnamefont {Li}}, \bibinfo {author} {\bibfnamefont {M.}~\bibnamefont
  {Coughlin}}, \bibinfo {author} {\bibfnamefont {M.~M.}\ \bibnamefont
  {Kozlov}}, \bibinfo {author} {\bibfnamefont {T.~A.}\ \bibnamefont
  {Rapoport}}, \ and\ \bibinfo {author} {\bibfnamefont {W.~A.}\ \bibnamefont
  {Prinz}},\ }\href {\doibase 10.1126/science.1153634} {\bibfield  {journal}
  {\bibinfo  {journal} {Science}\ }\textbf {\bibinfo {volume} {319}},\ \bibinfo
  {pages} {1247} (\bibinfo {year} {2008})},\ \Eprint
  {http://arxiv.org/abs/https://science.sciencemag.org/content/319/5867/1247.full.pdf}
  {https://science.sciencemag.org/content/319/5867/1247.full.pdf} \BibitemShut
  {NoStop}%
\bibitem [{\citenamefont {Shemesh}\ \emph {et~al.}(2014)\citenamefont
  {Shemesh}, \citenamefont {Klemm}, \citenamefont {Romano}, \citenamefont
  {Wang}, \citenamefont {Vaughan}, \citenamefont {Zhuang}, \citenamefont
  {Tukachinsky}, \citenamefont {Kozlov},\ and\ \citenamefont
  {Rapoport}}]{ShemeshER2014}%
  \BibitemOpen
  \bibfield  {author} {\bibinfo {author} {\bibfnamefont {T.}~\bibnamefont
  {Shemesh}}, \bibinfo {author} {\bibfnamefont {R.~W.}\ \bibnamefont {Klemm}},
  \bibinfo {author} {\bibfnamefont {F.~B.}\ \bibnamefont {Romano}}, \bibinfo
  {author} {\bibfnamefont {S.}~\bibnamefont {Wang}}, \bibinfo {author}
  {\bibfnamefont {J.}~\bibnamefont {Vaughan}}, \bibinfo {author} {\bibfnamefont
  {X.}~\bibnamefont {Zhuang}}, \bibinfo {author} {\bibfnamefont
  {H.}~\bibnamefont {Tukachinsky}}, \bibinfo {author} {\bibfnamefont {M.~M.}\
  \bibnamefont {Kozlov}}, \ and\ \bibinfo {author} {\bibfnamefont {T.~A.}\
  \bibnamefont {Rapoport}},\ }\href {\doibase 10.1073/pnas.1419997111}
  {\bibfield  {journal} {\bibinfo  {journal} {Proceedings of the National
  Academy of Sciences}\ }\textbf {\bibinfo {volume} {111}},\ \bibinfo {pages}
  {E5243} (\bibinfo {year} {2014})},\ \Eprint
  {http://arxiv.org/abs/https://www.pnas.org/content/111/49/E5243.full.pdf}
  {https://www.pnas.org/content/111/49/E5243.full.pdf} \BibitemShut {NoStop}%
\bibitem [{\citenamefont {Seifert}\ and\ \citenamefont
  {Lipowsky}(1995)}]{SeifertLipowsky1995}%
  \BibitemOpen
  \bibfield  {author} {\bibinfo {author} {\bibfnamefont {U.}~\bibnamefont
  {Seifert}}\ and\ \bibinfo {author} {\bibfnamefont {R.}~\bibnamefont
  {Lipowsky}},\ }\href@noop {} {\bibfield  {journal} {\bibinfo  {journal}
  {Structure and Dynamics of Membranes}\ }\textbf {\bibinfo {volume} {1}},\
  \bibinfo {pages} {403} (\bibinfo {year} {1995})}\BibitemShut {NoStop}%
\bibitem [{\citenamefont {Ramakrishnan}\ \emph {et~al.}(2018)\citenamefont
  {Ramakrishnan}, \citenamefont {Bradleh}, \citenamefont {Tourdot},\ and\
  \citenamefont {Radhakrishnan}}]{ReviewRamakrishnan2018}%
  \BibitemOpen
  \bibfield  {author} {\bibinfo {author} {\bibfnamefont {N.}~\bibnamefont
  {Ramakrishnan}}, \bibinfo {author} {\bibfnamefont {R.~P.}\ \bibnamefont
  {Bradleh}}, \bibinfo {author} {\bibfnamefont {R.~W.}\ \bibnamefont
  {Tourdot}}, \ and\ \bibinfo {author} {\bibfnamefont {R.}~\bibnamefont
  {Radhakrishnan}},\ }\href {\doibase https://doi.org/10.1088/1361-648X/aac703}
  {\bibfield  {journal} {\bibinfo  {journal} {J. Phys.: Cond. Matter}\ }\textbf
  {\bibinfo {volume} {30}},\ \bibinfo {pages} {273001} (\bibinfo {year}
  {2018})}\BibitemShut {NoStop}%
\bibitem [{\citenamefont {Lipowsky}(2021)}]{ReviewLipowsky2021}%
  \BibitemOpen
  \bibfield  {author} {\bibinfo {author} {\bibfnamefont {R.}~\bibnamefont
  {Lipowsky}},\ }\href {\doibase https://doi.org/10.1002/adbi.202101020}
  {\bibfield  {journal} {\bibinfo  {journal} {Advanced Biology}\ }\textbf
  {\bibinfo {volume} {6}},\ \bibinfo {pages} {2101020} (\bibinfo {year}
  {2021})}\BibitemShut {NoStop}%
\bibitem [{\citenamefont {Canham}(1970)}]{Canham1970}%
  \BibitemOpen
  \bibfield  {author} {\bibinfo {author} {\bibfnamefont {P.}~\bibnamefont
  {Canham}},\ }\href {\doibase https://doi.org/10.1016/S0022-5193(70)80032-7}
  {\bibfield  {journal} {\bibinfo  {journal} {Journal of Theoretical Biology}\
  }\textbf {\bibinfo {volume} {26}},\ \bibinfo {pages} {61 } (\bibinfo {year}
  {1970})}\BibitemShut {NoStop}%
\bibitem [{\citenamefont {Helfrich}(1973)}]{Helfrich1973}%
  \BibitemOpen
  \bibfield  {author} {\bibinfo {author} {\bibfnamefont {W.}~\bibnamefont
  {Helfrich}},\ }\href@noop {} {\bibfield  {journal} {\bibinfo  {journal}
  {Zeitschrift für Naturforschung}\ }\textbf {\bibinfo {volume} {28C}},\
  \bibinfo {pages} {693} (\bibinfo {year} {1973})}\BibitemShut {NoStop}%
\bibitem [{\citenamefont {Evans}(1974)}]{Evans1974}%
  \BibitemOpen
  \bibfield  {author} {\bibinfo {author} {\bibfnamefont {E.~A.}\ \bibnamefont
  {Evans}},\ }\href {\doibase https://doi.org/10.1016/S0006-3495(74)85959-X}
  {\bibfield  {journal} {\bibinfo  {journal} {Biophysical journal}\ }\textbf
  {\bibinfo {volume} {14}},\ \bibinfo {pages} {923–931} (\bibinfo {year}
  {1974})}\BibitemShut {NoStop}%
\bibitem [{\citenamefont {Seifert}\ \emph {et~al.}(1991)\citenamefont
  {Seifert}, \citenamefont {Berndl},\ and\ \citenamefont
  {Lipowsky}}]{SeifertLipowsky1991}%
  \BibitemOpen
  \bibfield  {author} {\bibinfo {author} {\bibfnamefont {U.}~\bibnamefont
  {Seifert}}, \bibinfo {author} {\bibfnamefont {K.}~\bibnamefont {Berndl}}, \
  and\ \bibinfo {author} {\bibfnamefont {R.}~\bibnamefont {Lipowsky}},\ }\href
  {\doibase 10.1103/PhysRevA.44.1182} {\bibfield  {journal} {\bibinfo
  {journal} {Phys. Rev. A}\ }\textbf {\bibinfo {volume} {44}},\ \bibinfo
  {pages} {1182} (\bibinfo {year} {1991})}\BibitemShut {NoStop}%
\bibitem [{\citenamefont {Gompper}\ and\ \citenamefont
  {Kroll}(1995)}]{Gompper1995}%
  \BibitemOpen
  \bibfield  {author} {\bibinfo {author} {\bibfnamefont {G.}~\bibnamefont
  {Gompper}}\ and\ \bibinfo {author} {\bibfnamefont {D.~M.}\ \bibnamefont
  {Kroll}},\ }\href {\doibase 10.1103/PhysRevE.51.514} {\bibfield  {journal}
  {\bibinfo  {journal} {Phys. Rev. E}\ }\textbf {\bibinfo {volume} {51}},\
  \bibinfo {pages} {514} (\bibinfo {year} {1995})}\BibitemShut {NoStop}%
\bibitem [{\citenamefont {Vanhille-Campos}\ and\ \citenamefont
  {Saric}(2021)}]{Vanhille-Campos2021}%
  \BibitemOpen
  \bibfield  {author} {\bibinfo {author} {\bibfnamefont {C.}~\bibnamefont
  {Vanhille-Campos}}\ and\ \bibinfo {author} {\bibfnamefont {A.}~\bibnamefont
  {Saric}},\ }\href {\doibase https://doi.org/10.1039/d0sm02012e} {\bibfield
  {journal} {\bibinfo  {journal} {Soft matter}\ }\textbf {\bibinfo {volume}
  {17}},\ \bibinfo {pages} {3798} (\bibinfo {year} {2021})}\BibitemShut
  {NoStop}%
\bibitem [{\citenamefont {Bo}\ and\ \citenamefont {Waugh}(1989)}]{Bo1989}%
  \BibitemOpen
  \bibfield  {author} {\bibinfo {author} {\bibfnamefont {L.}~\bibnamefont
  {Bo}}\ and\ \bibinfo {author} {\bibfnamefont {R.}~\bibnamefont {Waugh}},\
  }\href@noop {} {\bibfield  {journal} {\bibinfo  {journal} {Biophysical
  journal}\ }\textbf {\bibinfo {volume} {55 3}},\ \bibinfo {pages} {509}
  (\bibinfo {year} {1989})}\BibitemShut {NoStop}%
\bibitem [{\citenamefont {Bukman}\ \emph {et~al.}(1996)\citenamefont {Bukman},
  \citenamefont {Yao},\ and\ \citenamefont {Wortis}}]{Bukman1996}%
  \BibitemOpen
  \bibfield  {author} {\bibinfo {author} {\bibfnamefont {D.~J.}\ \bibnamefont
  {Bukman}}, \bibinfo {author} {\bibfnamefont {J.~H.}\ \bibnamefont {Yao}}, \
  and\ \bibinfo {author} {\bibfnamefont {M.}~\bibnamefont {Wortis}},\
  }\href@noop {} {\bibfield  {journal} {\bibinfo  {journal} {Phys. Rev. E}\
  }\textbf {\bibinfo {volume} {54}},\ \bibinfo {pages} {5463} (\bibinfo {year}
  {1996})}\BibitemShut {NoStop}%
\bibitem [{\citenamefont {Calladine}\ and\ \citenamefont
  {Greenwood}(2002)}]{Calladine2002}%
  \BibitemOpen
  \bibfield  {author} {\bibinfo {author} {\bibfnamefont {C.~R.}\ \bibnamefont
  {Calladine}}\ and\ \bibinfo {author} {\bibfnamefont {J.~A.}\ \bibnamefont
  {Greenwood}},\ }\href {\doibase https://doi.org/10.115/1.1500341} {\bibfield
  {journal} {\bibinfo  {journal} {J. of Biomech. Eng. -- Trans. of the ASME}\
  }\textbf {\bibinfo {volume} {124}},\ \bibinfo {pages} {576} (\bibinfo {year}
  {2002})}\BibitemShut {NoStop}%
\bibitem [{\citenamefont {Smith}\ \emph {et~al.}(2004)\citenamefont {Smith},
  \citenamefont {Sackmann},\ and\ \citenamefont {Seifert}}]{Seifert2004}%
  \BibitemOpen
  \bibfield  {author} {\bibinfo {author} {\bibfnamefont {A.-S. c.~v.}\
  \bibnamefont {Smith}}, \bibinfo {author} {\bibfnamefont {E.}~\bibnamefont
  {Sackmann}}, \ and\ \bibinfo {author} {\bibfnamefont {U.}~\bibnamefont
  {Seifert}},\ }\href {\doibase 10.1103/PhysRevLett.92.208101} {\bibfield
  {journal} {\bibinfo  {journal} {Phys. Rev. Lett.}\ }\textbf {\bibinfo
  {volume} {92}},\ \bibinfo {pages} {208101} (\bibinfo {year}
  {2004})}\BibitemShut {NoStop}%
\bibitem [{\citenamefont {Koster}\ \emph {et~al.}(2005)\citenamefont {Koster},
  \citenamefont {Cacciuto}, \citenamefont {Der\'enyi}, \citenamefont
  {Frenkel},\ and\ \citenamefont {Dogterom}}]{Koster2005}%
  \BibitemOpen
  \bibfield  {author} {\bibinfo {author} {\bibfnamefont {G.}~\bibnamefont
  {Koster}}, \bibinfo {author} {\bibfnamefont {A.}~\bibnamefont {Cacciuto}},
  \bibinfo {author} {\bibfnamefont {I.}~\bibnamefont {Der\'enyi}}, \bibinfo
  {author} {\bibfnamefont {D.}~\bibnamefont {Frenkel}}, \ and\ \bibinfo
  {author} {\bibfnamefont {M.}~\bibnamefont {Dogterom}},\ }\href {\doibase
  10.1103/PhysRevLett.94.068101} {\bibfield  {journal} {\bibinfo  {journal}
  {Phys. Rev. Lett.}\ }\textbf {\bibinfo {volume} {94}},\ \bibinfo {pages}
  {068101} (\bibinfo {year} {2005})}\BibitemShut {NoStop}%
\bibitem [{\citenamefont {Golushko}\ and\ \citenamefont
  {Rochal}(2016)}]{Golushko2016}%
  \BibitemOpen
  \bibfield  {author} {\bibinfo {author} {\bibfnamefont {I.}~\bibnamefont
  {Golushko}}\ and\ \bibinfo {author} {\bibfnamefont {S.}~\bibnamefont
  {Rochal}},\ }\href {\doibase 10.1134/S1063776116010027} {\bibfield  {journal}
  {\bibinfo  {journal} {Journal of Experimental and Theoretical Physics}\
  }\textbf {\bibinfo {volume} {122}},\ \bibinfo {pages} {169} (\bibinfo {year}
  {2016})}\BibitemShut {NoStop}%
\bibitem [{\citenamefont {Noguchi}(2021)}]{Noguchi2021}%
  \BibitemOpen
  \bibfield  {author} {\bibinfo {author} {\bibfnamefont {H.}~\bibnamefont
  {Noguchi}},\ }\href {\doibase https://doi.org/10.1039/d1sm01360b} {\bibfield
  {journal} {\bibinfo  {journal} {Soft Matter}\ }\textbf {\bibinfo {volume}
  {17}},\ \bibinfo {pages} {10469} (\bibinfo {year} {2021})}\BibitemShut
  {NoStop}%
\bibitem [{\citenamefont {Paraschif}\ \emph {et~al.}(2021)\citenamefont
  {Paraschif}, \citenamefont {Lagny}, \citenamefont {Vanhille-Campos},
  \citenamefont {Coudrier}, \citenamefont {Bassereau},\ and\ \citenamefont
  {Saric}}]{Paraschif2021}%
  \BibitemOpen
  \bibfield  {author} {\bibinfo {author} {\bibfnamefont {A.}~\bibnamefont
  {Paraschif}}, \bibinfo {author} {\bibfnamefont {T.~J.}\ \bibnamefont
  {Lagny}}, \bibinfo {author} {\bibfnamefont {C.}~\bibnamefont
  {Vanhille-Campos}}, \bibinfo {author} {\bibfnamefont {E.}~\bibnamefont
  {Coudrier}}, \bibinfo {author} {\bibfnamefont {P.}~\bibnamefont {Bassereau}},
  \ and\ \bibinfo {author} {\bibfnamefont {A.}~\bibnamefont {Saric}},\ }\href
  {\doibase https://doi.org/10.1016/j.bpj.2020.12.028} {\bibfield  {journal}
  {\bibinfo  {journal} {Biophysical Journal}\ }\textbf {\bibinfo {volume}
  {120}},\ \bibinfo {pages} {598} (\bibinfo {year} {2021})}\BibitemShut
  {NoStop}%
\bibitem [{\citenamefont {Bahrami}\ and\ \citenamefont
  {Hummer}(2017)}]{TriangulationBahrami2017}%
  \BibitemOpen
  \bibfield  {author} {\bibinfo {author} {\bibfnamefont {A.~H.}\ \bibnamefont
  {Bahrami}}\ and\ \bibinfo {author} {\bibfnamefont {G.}~\bibnamefont
  {Hummer}},\ }\href {\doibase 10.1021/acsnano.7b05542} {\bibfield  {journal}
  {\bibinfo  {journal} {ACS Nano}\ }\textbf {\bibinfo {volume} {11}},\ \bibinfo
  {pages} {9558} (\bibinfo {year} {2017})}\BibitemShut {NoStop}%
\bibitem [{\citenamefont {Rawicz}\ \emph {et~al.}(2000)\citenamefont {Rawicz},
  \citenamefont {Olbrich}, \citenamefont {McIntosh}, \citenamefont {Needham},\
  and\ \citenamefont {Evans}}]{RAWICZ2000}%
  \BibitemOpen
  \bibfield  {author} {\bibinfo {author} {\bibfnamefont {W.}~\bibnamefont
  {Rawicz}}, \bibinfo {author} {\bibfnamefont {K.}~\bibnamefont {Olbrich}},
  \bibinfo {author} {\bibfnamefont {T.}~\bibnamefont {McIntosh}}, \bibinfo
  {author} {\bibfnamefont {D.}~\bibnamefont {Needham}}, \ and\ \bibinfo
  {author} {\bibfnamefont {E.}~\bibnamefont {Evans}},\ }\href {\doibase
  https://doi.org/10.1016/S0006-3495(00)76295-3} {\bibfield  {journal}
  {\bibinfo  {journal} {Biophysical Journal}\ }\textbf {\bibinfo {volume}
  {79}},\ \bibinfo {pages} {328 } (\bibinfo {year} {2000})}\BibitemShut
  {NoStop}%
\bibitem [{\citenamefont {Carmo}(1976)}]{DoCarmoDiffGeo1976}%
  \BibitemOpen
  \bibfield  {author} {\bibinfo {author} {\bibfnamefont {M.~D.}\ \bibnamefont
  {Carmo}},\ }\href@noop {} {\emph {\bibinfo {title} {Differential Geometry of
  Curves and Surfaces}}}\ (\bibinfo  {publisher} {Prentice Hall},\ \bibinfo
  {year} {1976})\BibitemShut {NoStop}%
\bibitem [{\citenamefont {Bozic}\ \emph {et~al.}(1992)\citenamefont {Bozic},
  \citenamefont {Svetina}, \citenamefont {Zeks},\ and\ \citenamefont
  {Waugh}}]{ADE_Bozic1992}%
  \BibitemOpen
  \bibfield  {author} {\bibinfo {author} {\bibfnamefont {B.}~\bibnamefont
  {Bozic}}, \bibinfo {author} {\bibfnamefont {S.}~\bibnamefont {Svetina}},
  \bibinfo {author} {\bibfnamefont {B.}~\bibnamefont {Zeks}}, \ and\ \bibinfo
  {author} {\bibfnamefont {R.}~\bibnamefont {Waugh}},\ }\href {\doibase
  10.1016/s0006-3495(92)81903-3} {\bibfield  {journal} {\bibinfo  {journal}
  {Biophysical journal}\ }\textbf {\bibinfo {volume} {61}},\ \bibinfo {pages}
  {963—973} (\bibinfo {year} {1992})}\BibitemShut {NoStop}%
\bibitem [{\citenamefont {Wiese}\ \emph {et~al.}(1992)\citenamefont {Wiese},
  \citenamefont {Harbich},\ and\ \citenamefont {Helfrich}}]{ADE_Wiese1992}%
  \BibitemOpen
  \bibfield  {author} {\bibinfo {author} {\bibfnamefont {W.}~\bibnamefont
  {Wiese}}, \bibinfo {author} {\bibfnamefont {W.}~\bibnamefont {Harbich}}, \
  and\ \bibinfo {author} {\bibfnamefont {W.}~\bibnamefont {Helfrich}},\ }\href
  {\doibase 10.1088/0953-8984/4/7/004} {\bibfield  {journal} {\bibinfo
  {journal} {Journal of Physics: Condensed Matter}\ }\textbf {\bibinfo {volume}
  {4}},\ \bibinfo {pages} {1647} (\bibinfo {year} {1992})}\BibitemShut
  {NoStop}%
\bibitem [{\citenamefont {Heinrich}\ \emph {et~al.}(1993)\citenamefont
  {Heinrich}, \citenamefont {Svetina},\ and\ \citenamefont {\ifmmode
  \check{Z}\else \v{Z}\fi{}ek\ifmmode~\check{s}\else
  \v{s}\fi{}}}]{ADE_Heinrich_1993}%
  \BibitemOpen
  \bibfield  {author} {\bibinfo {author} {\bibfnamefont {V.}~\bibnamefont
  {Heinrich}}, \bibinfo {author} {\bibfnamefont {S.~c.~v.}\ \bibnamefont
  {Svetina}}, \ and\ \bibinfo {author} {\bibfnamefont {B.~c.~v.}\ \bibnamefont
  {\ifmmode \check{Z}\else \v{Z}\fi{}ek\ifmmode~\check{s}\else \v{s}\fi{}}},\
  }\href {\doibase 10.1103/PhysRevE.48.3112} {\bibfield  {journal} {\bibinfo
  {journal} {Phys. Rev. E}\ }\textbf {\bibinfo {volume} {48}},\ \bibinfo
  {pages} {3112} (\bibinfo {year} {1993})}\BibitemShut {NoStop}%
\bibitem [{\citenamefont {Miao}\ \emph {et~al.}(1994)\citenamefont {Miao},
  \citenamefont {Seifert}, \citenamefont {Wortis},\ and\ \citenamefont
  {D\"obereiner}}]{ADE_Miao_1994}%
  \BibitemOpen
  \bibfield  {author} {\bibinfo {author} {\bibfnamefont {L.}~\bibnamefont
  {Miao}}, \bibinfo {author} {\bibfnamefont {U.}~\bibnamefont {Seifert}},
  \bibinfo {author} {\bibfnamefont {M.}~\bibnamefont {Wortis}}, \ and\ \bibinfo
  {author} {\bibfnamefont {H.-G.}\ \bibnamefont {D\"obereiner}},\ }\href
  {\doibase 10.1103/PhysRevE.49.5389} {\bibfield  {journal} {\bibinfo
  {journal} {Phys. Rev. E}\ }\textbf {\bibinfo {volume} {49}},\ \bibinfo
  {pages} {5389} (\bibinfo {year} {1994})}\BibitemShut {NoStop}%
\bibitem [{\citenamefont {Sheetz}\ and\ \citenamefont
  {Singer}(1974)}]{BC_1974}%
  \BibitemOpen
  \bibfield  {author} {\bibinfo {author} {\bibfnamefont {M.}~\bibnamefont
  {Sheetz}}\ and\ \bibinfo {author} {\bibfnamefont {S.}~\bibnamefont
  {Singer}},\ }\href {\doibase 10.1073/pnas.71.11.4457} {\bibfield  {journal}
  {\bibinfo  {journal} {Proceedings of the National Academy of Sciences of the
  United States of America}\ }\textbf {\bibinfo {volume} {71}},\ \bibinfo
  {pages} {4457—4461} (\bibinfo {year} {1974})}\BibitemShut {NoStop}%
\bibitem [{\citenamefont {Svetina}\ and\ \citenamefont {Zeks}(1989)}]{BC_1989}%
  \BibitemOpen
  \bibfield  {author} {\bibinfo {author} {\bibfnamefont {S.}~\bibnamefont
  {Svetina}}\ and\ \bibinfo {author} {\bibfnamefont {B.}~\bibnamefont {Zeks}},\
  }\href {\doibase 10.1007/bf00257107} {\bibfield  {journal} {\bibinfo
  {journal} {European biophysics journal : EBJ}\ }\textbf {\bibinfo {volume}
  {17}},\ \bibinfo {pages} {101—111} (\bibinfo {year} {1989})}\BibitemShut
  {NoStop}%
\bibitem [{\citenamefont {Ziherl}\ and\ \citenamefont
  {Svetina}(2005)}]{Ziherl2005}%
  \BibitemOpen
  \bibfield  {author} {\bibinfo {author} {\bibfnamefont {P.}~\bibnamefont
  {Ziherl}}\ and\ \bibinfo {author} {\bibfnamefont {S.}~\bibnamefont
  {Svetina}},\ }\href {\doibase 10.1209/epl/i2004-10527-4} {\bibfield
  {journal} {\bibinfo  {journal} {Europhysics Letters ({EPL})}\ }\textbf
  {\bibinfo {volume} {70}},\ \bibinfo {pages} {690} (\bibinfo {year}
  {2005})}\BibitemShut {NoStop}%
\bibitem [{\citenamefont {Itzykson}(1986)}]{TriangulationItzykson1986}%
  \BibitemOpen
  \bibfield  {author} {\bibinfo {author} {\bibnamefont {Itzykson}},\
  }\href@noop {} {\bibfield  {journal} {\bibinfo  {journal} {Proceedings of the
  GIFT seminar, Jaca85, WorldScientific, Singapore}\ } (\bibinfo {year}
  {1986})}\BibitemShut {NoStop}%
\bibitem [{\citenamefont {Kantor}\ and\ \citenamefont
  {Nelson}(1987)}]{TriangulationKantor1987}%
  \BibitemOpen
  \bibfield  {author} {\bibinfo {author} {\bibfnamefont {Y.}~\bibnamefont
  {Kantor}}\ and\ \bibinfo {author} {\bibfnamefont {D.~R.}\ \bibnamefont
  {Nelson}},\ }\href {\doibase 10.1103/PhysRevLett.58.2774} {\bibfield
  {journal} {\bibinfo  {journal} {Phys. Rev. Lett.}\ }\textbf {\bibinfo
  {volume} {58}},\ \bibinfo {pages} {2774} (\bibinfo {year}
  {1987})}\BibitemShut {NoStop}%
\bibitem [{\citenamefont {{G. Gompper}}\ and\ \citenamefont {{D.M.
  Kroll}}(1996)}]{TriangulationGompper1996}%
  \BibitemOpen
  \bibfield  {author} {\bibinfo {author} {\bibnamefont {{G. Gompper}}}\ and\
  \bibinfo {author} {\bibnamefont {{D.M. Kroll}}},\ }\href {\doibase
  10.1051/jp1:1996246} {\bibfield  {journal} {\bibinfo  {journal} {J. Phys. I
  France}\ }\textbf {\bibinfo {volume} {6}},\ \bibinfo {pages} {1305} (\bibinfo
  {year} {1996})}\BibitemShut {NoStop}%
\bibitem [{\citenamefont {Gompper}\ and\ \citenamefont
  {Kroll}(1997)}]{TriangulationGompper1997}%
  \BibitemOpen
  \bibfield  {author} {\bibinfo {author} {\bibfnamefont {G.}~\bibnamefont
  {Gompper}}\ and\ \bibinfo {author} {\bibfnamefont {D.~M.}\ \bibnamefont
  {Kroll}},\ }\href {\doibase 10.1088/0953-8984/9/42/001} {\bibfield  {journal}
  {\bibinfo  {journal} {Journal of Physics: Condensed Matter}\ }\textbf
  {\bibinfo {volume} {9}},\ \bibinfo {pages} {8795} (\bibinfo {year}
  {1997})}\BibitemShut {NoStop}%
\bibitem [{\citenamefont {J{\"u}licher}(1996)}]{TriangulationJulicher1996}%
  \BibitemOpen
  \bibfield  {author} {\bibinfo {author} {\bibfnamefont {F.}~\bibnamefont
  {J{\"u}licher}},\ }\href@noop {} {\bibfield  {journal} {\bibinfo  {journal}
  {Journal de Physique II}\ }\textbf {\bibinfo {volume} {6}},\ \bibinfo {pages}
  {1797} (\bibinfo {year} {1996})}\BibitemShut {NoStop}%
\bibitem [{\citenamefont {\ifmmode \check{S}\else
  \v{S}\fi{}ari\ifmmode~\acute{c}\else \'{c}\fi{}}\ and\ \citenamefont
  {Cacciuto}(2012)}]{TriangulationSaric2012}%
  \BibitemOpen
  \bibfield  {author} {\bibinfo {author} {\bibfnamefont {A.}~\bibnamefont
  {\ifmmode \check{S}\else \v{S}\fi{}ari\ifmmode~\acute{c}\else \'{c}\fi{}}} \
  and\ \bibinfo {author} {\bibfnamefont {A.}~\bibnamefont {Cacciuto}},\ }\href
  {\doibase 10.1103/PhysRevLett.109.188101} {\bibfield  {journal} {\bibinfo
  {journal} {Phys. Rev. Lett.}\ }\textbf {\bibinfo {volume} {109}},\ \bibinfo
  {pages} {188101} (\bibinfo {year} {2012})}\BibitemShut {NoStop}%
\bibitem [{\citenamefont {Bahrami}\ \emph {et~al.}(2012)\citenamefont
  {Bahrami}, \citenamefont {Lipowsky},\ and\ \citenamefont
  {Weikl}}]{TriangulationBahrami2012}%
  \BibitemOpen
  \bibfield  {author} {\bibinfo {author} {\bibfnamefont {A.~H.}\ \bibnamefont
  {Bahrami}}, \bibinfo {author} {\bibfnamefont {R.}~\bibnamefont {Lipowsky}}, \
  and\ \bibinfo {author} {\bibfnamefont {T.~R.}\ \bibnamefont {Weikl}},\ }\href
  {\doibase 10.1103/PhysRevLett.109.188102} {\bibfield  {journal} {\bibinfo
  {journal} {Phys. Rev. Lett.}\ }\textbf {\bibinfo {volume} {109}},\ \bibinfo
  {pages} {188102} (\bibinfo {year} {2012})}\BibitemShut {NoStop}%
\bibitem [{\citenamefont {Ramakrishnan}\ \emph {et~al.}(2013)\citenamefont
  {Ramakrishnan}, \citenamefont {Sunil~Kumar},\ and\ \citenamefont
  {Ipsen}}]{TriangulationRamakrishnan2013}%
  \BibitemOpen
  \bibfield  {author} {\bibinfo {author} {\bibfnamefont {N.}~\bibnamefont
  {Ramakrishnan}}, \bibinfo {author} {\bibfnamefont {P.}~\bibnamefont
  {Sunil~Kumar}}, \ and\ \bibinfo {author} {\bibfnamefont {J.~H.}\ \bibnamefont
  {Ipsen}},\ }\href {\doibase 10.1016/j.bpj.2012.12.045} {\bibfield  {journal}
  {\bibinfo  {journal} {Biophysical Journal}\ }\textbf {\bibinfo {volume}
  {104}},\ \bibinfo {pages} {1018–1028} (\bibinfo {year} {2013})}\BibitemShut
  {NoStop}%
\bibitem [{\citenamefont {Vahid}\ \emph {et~al.}(2017)\citenamefont {Vahid},
  \citenamefont {Šarić},\ and\ \citenamefont
  {Idema}}]{TriangulationVahid2017}%
  \BibitemOpen
  \bibfield  {author} {\bibinfo {author} {\bibfnamefont {A.}~\bibnamefont
  {Vahid}}, \bibinfo {author} {\bibfnamefont {A.}~\bibnamefont {Šarić}}, \
  and\ \bibinfo {author} {\bibfnamefont {T.}~\bibnamefont {Idema}},\ }\href
  {\doibase 10.1039/C7SM00433H} {\bibfield  {journal} {\bibinfo  {journal}
  {Soft Matter}\ }\textbf {\bibinfo {volume} {13}},\ \bibinfo {pages} {4924}
  (\bibinfo {year} {2017})}\BibitemShut {NoStop}%
\bibitem [{\citenamefont {Li}\ and\ \citenamefont
  {Abel}(2018)}]{TriangulatedLi2018}%
  \BibitemOpen
  \bibfield  {author} {\bibinfo {author} {\bibfnamefont {B.}~\bibnamefont
  {Li}}\ and\ \bibinfo {author} {\bibfnamefont {S.~M.}\ \bibnamefont {Abel}},\
  }\href {\doibase 10.1039/C7SM01751K} {\bibfield  {journal} {\bibinfo
  {journal} {Soft Matter}\ }\textbf {\bibinfo {volume} {14}},\ \bibinfo {pages}
  {185} (\bibinfo {year} {2018})}\BibitemShut {NoStop}%
\bibitem [{\citenamefont {Hoore}\ \emph {et~al.}(2018)\citenamefont {Hoore},
  \citenamefont {Yaya}, \citenamefont {Podgorski}, \citenamefont {Wagner},
  \citenamefont {Gompper},\ and\ \citenamefont
  {Fedosov}}]{TriangulationHoore2018}%
  \BibitemOpen
  \bibfield  {author} {\bibinfo {author} {\bibfnamefont {M.}~\bibnamefont
  {Hoore}}, \bibinfo {author} {\bibfnamefont {F.}~\bibnamefont {Yaya}},
  \bibinfo {author} {\bibfnamefont {T.}~\bibnamefont {Podgorski}}, \bibinfo
  {author} {\bibfnamefont {C.}~\bibnamefont {Wagner}}, \bibinfo {author}
  {\bibfnamefont {G.}~\bibnamefont {Gompper}}, \ and\ \bibinfo {author}
  {\bibfnamefont {D.~A.}\ \bibnamefont {Fedosov}},\ }\href {\doibase
  10.1039/C8SM00634B} {\bibfield  {journal} {\bibinfo  {journal} {Soft Matter}\
  }\textbf {\bibinfo {volume} {14}},\ \bibinfo {pages} {6278} (\bibinfo {year}
  {2018})}\BibitemShut {NoStop}%
\bibitem [{\citenamefont {Bian}\ \emph {et~al.}(2020)\citenamefont {Bian},
  \citenamefont {Litvinov},\ and\ \citenamefont {Koumoutsakos}}]{BIAN2020}%
  \BibitemOpen
  \bibfield  {author} {\bibinfo {author} {\bibfnamefont {X.}~\bibnamefont
  {Bian}}, \bibinfo {author} {\bibfnamefont {S.}~\bibnamefont {Litvinov}}, \
  and\ \bibinfo {author} {\bibfnamefont {P.}~\bibnamefont {Koumoutsakos}},\
  }\href {\doibase https://doi.org/10.1016/j.cma.2019.112758} {\bibfield
  {journal} {\bibinfo  {journal} {Computer Methods in Applied Mechanics and
  Engineering}\ }\textbf {\bibinfo {volume} {359}},\ \bibinfo {pages} {112758}
  (\bibinfo {year} {2020})}\BibitemShut {NoStop}%
\bibitem [{\citenamefont {Noguchi}\ and\ \citenamefont
  {Gompper}(2004)}]{TriangulationNoguchi2004}%
  \BibitemOpen
  \bibfield  {author} {\bibinfo {author} {\bibfnamefont {H.}~\bibnamefont
  {Noguchi}}\ and\ \bibinfo {author} {\bibfnamefont {G.}~\bibnamefont
  {Gompper}},\ }\href {\doibase 10.1103/PhysRevLett.93.258102} {\bibfield
  {journal} {\bibinfo  {journal} {Phys. Rev. Lett.}\ }\textbf {\bibinfo
  {volume} {93}},\ \bibinfo {pages} {258102} (\bibinfo {year}
  {2004})}\BibitemShut {NoStop}%
\bibitem [{\citenamefont {Noguchi}\ and\ \citenamefont
  {Gompper}(2005)}]{TriangulationNoguchi2005}%
  \BibitemOpen
  \bibfield  {author} {\bibinfo {author} {\bibfnamefont {H.}~\bibnamefont
  {Noguchi}}\ and\ \bibinfo {author} {\bibfnamefont {G.}~\bibnamefont
  {Gompper}},\ }\href {\doibase 10.1103/PhysRevE.72.011901} {\bibfield
  {journal} {\bibinfo  {journal} {Phys. Rev. E}\ }\textbf {\bibinfo {volume}
  {72}},\ \bibinfo {pages} {011901} (\bibinfo {year} {2005})}\BibitemShut
  {NoStop}%
\bibitem [{\citenamefont {Lobovkina}\ \emph {et~al.}(2004)\citenamefont
  {Lobovkina}, \citenamefont {Dommersnes}, \citenamefont {Joanny},
  \citenamefont {Bassereau}, \citenamefont {Karlsson},\ and\ \citenamefont
  {Orwar}}]{Lobovkina2004}%
  \BibitemOpen
  \bibfield  {author} {\bibinfo {author} {\bibfnamefont {T.}~\bibnamefont
  {Lobovkina}}, \bibinfo {author} {\bibfnamefont {P.}~\bibnamefont
  {Dommersnes}}, \bibinfo {author} {\bibfnamefont {J.~F.}\ \bibnamefont
  {Joanny}}, \bibinfo {author} {\bibfnamefont {P.}~\bibnamefont {Bassereau}},
  \bibinfo {author} {\bibfnamefont {M.}~\bibnamefont {Karlsson}}, \ and\
  \bibinfo {author} {\bibfnamefont {O.}~\bibnamefont {Orwar}},\ }\href@noop {}
  {\bibfield  {journal} {\bibinfo  {journal} {Proceedings of the National
  Academy of Sciences of the United States of America}\ }\textbf {\bibinfo
  {volume} {101 21}},\ \bibinfo {pages} {7949} (\bibinfo {year}
  {2004})}\BibitemShut {NoStop}%
\bibitem [{\citenamefont {Lobovkina}\ \emph {et~al.}(2006)\citenamefont
  {Lobovkina}, \citenamefont {Dommersnes}, \citenamefont {Joanny},
  \citenamefont {Hurtig},\ and\ \citenamefont {Orwar}}]{Lobovkina2006}%
  \BibitemOpen
  \bibfield  {author} {\bibinfo {author} {\bibfnamefont {T.}~\bibnamefont
  {Lobovkina}}, \bibinfo {author} {\bibfnamefont {P.}~\bibnamefont
  {Dommersnes}}, \bibinfo {author} {\bibfnamefont {J.-F.}\ \bibnamefont
  {Joanny}}, \bibinfo {author} {\bibfnamefont {J.}~\bibnamefont {Hurtig}}, \
  and\ \bibinfo {author} {\bibfnamefont {O.}~\bibnamefont {Orwar}},\ }\href
  {\doibase 10.1103/PhysRevLett.97.188105} {\bibfield  {journal} {\bibinfo
  {journal} {Phys. Rev. Lett.}\ }\textbf {\bibinfo {volume} {97}},\ \bibinfo
  {pages} {188105} (\bibinfo {year} {2006})}\BibitemShut {NoStop}%
\bibitem [{\citenamefont {Lobovkina}\ \emph {et~al.}(2008)\citenamefont
  {Lobovkina}, \citenamefont {Dommersnes}, \citenamefont {Tiourine},
  \citenamefont {Joanny},\ and\ \citenamefont {Orwar}}]{Lobovkina2008}%
  \BibitemOpen
  \bibfield  {author} {\bibinfo {author} {\bibfnamefont {T.}~\bibnamefont
  {Lobovkina}}, \bibinfo {author} {\bibfnamefont {P.}~\bibnamefont
  {Dommersnes}}, \bibinfo {author} {\bibfnamefont {S.}~\bibnamefont
  {Tiourine}}, \bibinfo {author} {\bibfnamefont {J.}~\bibnamefont {Joanny}}, \
  and\ \bibinfo {author} {\bibfnamefont {O.}~\bibnamefont {Orwar}},\ }\href
  {\doibase 10.1140/epje/i2007-10325-x} {\bibfield  {journal} {\bibinfo
  {journal} {The European physical journal. E, Soft matter}\ }\textbf {\bibinfo
  {volume} {26}},\ \bibinfo {pages} {295} (\bibinfo {year} {2008})}\BibitemShut
  {NoStop}%
 \bibitem{note} A small
     compensating force is applied to all $N$ vertices to ensure
     that the total force is still zero.%
\bibitem [{\citenamefont {Stillinger}\ and\ \citenamefont
   {Weber}(1985)}]{StillingerWeber1985}%
   \BibitemOpen
   \bibfield  {author} {\bibinfo {author} {\bibfnamefont {F.~H.}\ \bibnamefont
   {Stillinger}}\ and\ \bibinfo {author} {\bibfnamefont {T.~A.}\ \bibnamefont
   {Weber}},\ }\href {\doibase 10.1103/PhysRevB.31.5262} {\bibfield  {journal}
   {\bibinfo  {journal} {Phys. Rev. B}\ }\textbf {\bibinfo {volume} {31}},\
   \bibinfo {pages} {5262} (\bibinfo {year} {1985})}\BibitemShut {NoStop}%
\bibitem [{Note1()}]{Note1}%
   \BibitemOpen
   \bibinfo {note} {The area difference $\Delta A$ is usually given by $\Delta A
   = 2 h \DOTSI \intop \ilimits@ \protect \text {d} A K $, where $h$ is the
   thickness of the membrane\cite {BIAN2020}. However, different conventions
   exist\cite
   {BC_1974,Evans1974,BC_1989,ADE_Bozic1992,ADE_Wiese1992,ADE_Heinrich_1993,ADE_Miao_1994,TriangulationBahrami2017,BIAN2020},
   which makes it more convenient to use the renormalized area difference
   $\Delta a$ that is one for a sphere.}\BibitemShut {Stop}%
\end{thebibliography}


%

\newpage
\newpage

\begin{center}
{\large \bf Supplementary Information}
\end{center}

\subsection{Dynamically-triangulated membrane model}

For the simulations the dynamically-triangulated membrane model of
Ref.\ \citen{TriangulationNoguchi2005} was used. The model is
described in detail in that reference.  Here we briefly recapitulate
the potentials and parameters for the convenience of the reader.

\subsubsection*{Curvature Energy}

The shape of the vesicle is mostly controlled by the curvature energy
given by \revision{Eq.~(1) in the main text}.  This equation is 
discretized as follows
\cite{TriangulationGompper1996,TriangulationItzykson1986}:
\begin{equation}
U_{\text{cv}} =  \frac{\kappa}{2}\sum_i \frac{1}{\sigma_i}\left( \sum_{j(i)} 
\frac{\sigma_{i,j} \vec{r}_{i,j}}{r_{i,j}}  \right)^2
\end{equation}
The value for the bending rigidity for lipid membranes is typically
$\kappa = 20\,k_{\text{B}}T$ \cite{RAWICZ2000}, where $k_{\text{B}}T$
is the thermal energy. The first sum goes over all vertices $i$ and
the second sum goes over all neighbors of the vertex $i$, $j(i)$, that
are connected by bonds. The vector between vertices $i$ and $j$ is
denoted by $\vec{r}_{i,j}$ and $r_{i,j} = \| \vec{r}_{i,j} \|$.
$\sigma_{i,j}$ is the length of the bond in the dual lattice, which is
given by $\sigma_{i,j} = r_{i,j} [ \cot(\theta_1) + \cot(\theta_2) ]/2
$, where $\theta_1$ and $\theta_2$ are the angles opposite to the bond
connecting $i$ and $j$ in the two triangles sharing this bond. The
area of the dual cell of vertex $i$ is given by $ \sigma_i = 0.25
\sum_{j(i)} \sigma_{i,j} r_{i,j} $.

\subsubsection*{Bond and Repulsive Interactions}

In order to perform molecular dynamics simulations a Stillinger-Weber
potential \cite{StillingerWeber1985} is used to describe bond and
excluded-volume interactions between vertices.  All vertices connected
by bonds interact via the following attractive potential:
\begin{equation}
U_{\text{bond}}(r_{i,j}) = \varepsilon
\begin{dcases}
\frac{l_b \exp[l_b/(l_{\text{c0}} - r_{i,j})]}{l_{\text{max}}- r_{i,j}} & 
(r_{i,j} > l_{\text{c0}})\\
0              & (r_{i,j} \leq l_{\text{c0}}).
\end{dcases}
\end{equation}
At short distances all particles interact via the following repulsive 
excluded volume potential:
\begin{equation}
U_{\text{rep}}(r_{i,j}) = \varepsilon
\begin{dcases}
\frac{l_b \exp[l_b/(r_{i,j} - l_{\text{c1}})]}{r_{i,j} - l_{\text{min}} } & 
(r_{i,j} < l_{\text{c1}})\\
0              & (r_{i,j} \geq l_{\text{c1}})
\end{dcases}
\end{equation}
The parameters of the potentials are listed in Table
\ref{tab:parameters_potential}, where the parameter $l_b$ refers to
the bond length and the parameter $\varepsilon=80 k_\text{B}T$ is a
constant energy prefactor.  This bond length is the length unit and
hence set to one in all simulations. 

\begin{table}[H]
	\centering
	\begin{tabular}[t]{lcc}
		
		\hline
		Parameter &Value &Description\\
		\hline
		$l_{\text{max}}$ & $1.33 l_b$ & maximum bond length\\
		$l_{\text{min}}$ & $0.67 l_b$ & minimum distance between two vertices\\
		\hline
		$l_{\text{c0}}$ & $1.15 l_b$ & cutoff length for $U_{\text{bond}}$\\
		$l_{\text{c1}}$ & $0.85 l_b$ & cutoff length for $U_{\text{rep}}$\\
		\hline
		
	\end{tabular}
	\caption{Parameters used for the bond and repulsive interactions.}
	\label{tab:parameters_potential}
\end{table}

\subsubsection*{Area and Volume}

The total area is the sum over the area of each vertex $A_i$, which is
given by the weighted sum over the area of all neighboring triangles $
A_\alpha $:
\begin{equation}
A = \sum_{i=1}^N A_i \text{\ \ \ \ with \ \ \ \ } A_i = \frac{1}{3}\sum_{\alpha\in \text{neigh. triangles}} A_\alpha
\end{equation}
Here and in the following quantities with greek indices denote
triangles and roman indices denote vertices.

The volume $V$ enclosed by the membrane is calculated
as \cite{TriangulationBahrami2017}:
\begin{equation}
V = \sum_{\alpha=1}^{N_{\text{t}}} V_\alpha 
\text{\ \ with signed subvolumes \ \  } 
V_\alpha = \frac{1}{3} (\hat{n}_\alpha \cdot \vec{R}_\alpha) A_\alpha,
\end{equation}
where $\hat{n}_\alpha$ is the unit normal vector of triangle $\alpha$
pointing outwards and $\vec{R}_\alpha$ is the position vector of one
of the vertices of the triangle relative to an a reference point. This
reference point can, in fact, be chosen arbitrarily and can even lie
outside of the structure, because any additional contribution from
outside the vesicle will eventually be subtracted by another
subvolume $V_{\alpha}$.

\subsubsection*{Area Difference}

The renormalized area difference \footnote{The area difference $\Delta
A$ is usually given by $\Delta A = 2 h \int \text{d} A K $, where $h$
is the thickness of the membrane\cite{BIAN2020}. However, different
conventions
exist\cite{BC_1974,Evans1974,BC_1989,ADE_Bozic1992,ADE_Wiese1992,ADE_Heinrich_1993,ADE_Miao_1994,TriangulationBahrami2017,BIAN2020},
which makes it more convenient to use the renormalized area difference
$\Delta a$ that is one for a sphere.} $\Delta a$ is defined as:
\begin{equation}
\label{eq:DAvalue}
\Delta a =  \frac{1}{4 \sqrt{\pi A_0}} \int \text{d} A K = \frac{1}{4 \sqrt{\pi A_0}}\sum_{i=1}^{N} |\vec{H}_i| \, \frac{\vec{H}_i \cdot \hat{n}_i}{|\vec{H}_i \cdot \hat{n}_i|}
\end{equation}
where $\vec{H}_i$ is the oriented curvature contribution of vertex $i$,
\begin{equation}
\vec{H}_i = \sum_{j(i)} 
\frac{\sigma_{i,j} \vec{r}_{i,j}}{r_{i,j}}.
\end{equation}
Here, the term ${\vec{H}_i \cdot \hat{n}_i}/{|\vec{H}_i \cdot
\hat{n}_i|} $  gives the orientation of the curvature, i.e. if it is
convex (+1) or concave (-1) using the surface normal vector $
\hat{n}_i $, as the average orientation of the neighboring triangles.
The normalization is chosen such that a sphere has an area difference
of $ \Delta a = 1 $. 

\vspace*{-\baselineskip}

\revision{
\subsubsection*{Constraint Potentials}}

\revision{
In our simulations, the area $A$ of branched and linear structures
is constrained to $A = A_0$ by introducing a constraint potential
\begin{equation}
U_{\text{A}} = \frac{1}{2} k_{\text{A}} (A - A_0)^2.
\end{equation}
In some simulations, additional constraint potentials 
$U_{\text{V}}$ and/or $U_{\Delta a}$ are included to fix the 
enclosed volume at $V = V_0$ via 
\begin{equation}
U_{\text{V}} = \frac{1}{2} k_{\text{V}} (V - V_0)^2
\end{equation}
and/or the renormalized area difference $\Delta a$ at
$\Delta a_0$ via
\begin{equation} 
 U_{\Delta a} =
\frac{1}{2} k_{\Delta a} (\Delta a - \Delta a_0)^2.
\end{equation}
Since the constraints are implemented by harmonic potentials
rather than being strictly enforced, small variations in $A$, $V$, 
and $\Delta a$ are still possible even in the presence of constraint
potentials. For example, in Table 1 of the main text, the
actual values sometimes slightly differ (by less than 1\%)
from the imposed values.
}

\bigskip

\subsubsection*{Overall Potential}

Finally, the overall potential used in the simulations is a
combination of all the potentials described above:
\begin{equation}
U_{\text{tot}} = U_{\text{cv}} + U_{\text{bond}} + U_{\text{rep}} + 
U_{\text{A}} + U_{\text{V}} + U_{\Delta a}
\label{eq:total_potential}
\end{equation}
The total Hamiltonian of the system is therefore:
\begin{equation}
\label{eq:Hnew}
H_0 = \sum_{i=1}^{N} \frac{\vec{p}_i^2}{2m} + U_{\text{tot}}
\end{equation}
where $\vec{p}_i$ are the momenta of the vertices and $m$ their masses.

\subsection{Time evolution, transformation pathways, and movies}

\subsubsection{Higher order junctions subject to external forces}

\begin{figure}[h!]	
		\centering
		\includegraphics[width=0.9\linewidth]
        {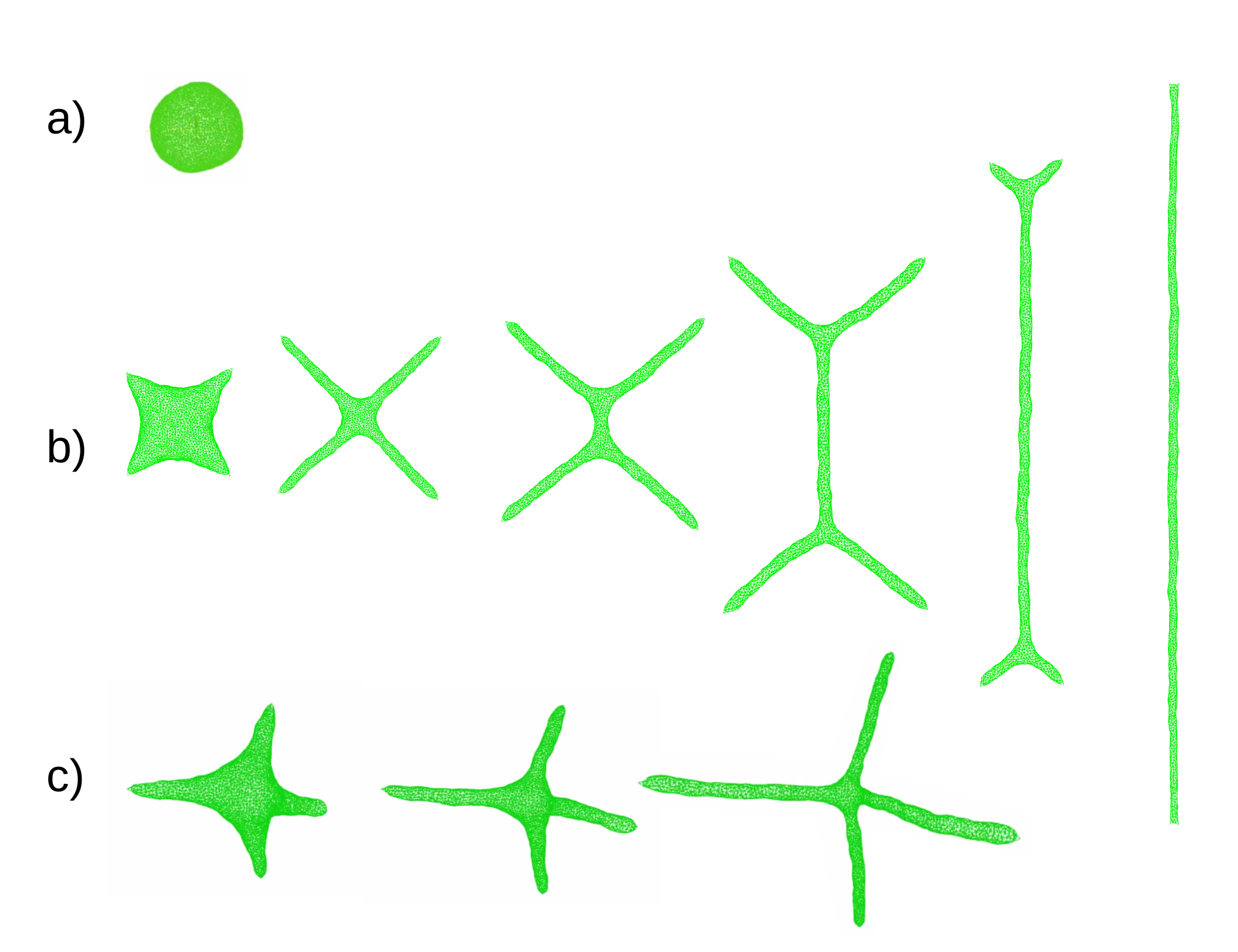}
	\caption{
    Time evolution of mechanically enforced tubular structures
    with intermediate higher order junctions.
    (a) Initial state: A small sphere (no volume constraint).
    (b) Evolution after applying four forces $F_{\text{ext}}$ 
        in a plane with angle $90^\circ$. First a four-fold
        junction forms. Then this junction separates into two
        Y-junctions. Since they do not have the correct angles
        $120^\circ$, they move outwards. The final stable state is a tube.
        The same behavior is observed if the forces on the branches 
        are twisted with respect to each other (not shown).
   (c) Evolution if the four forces $F_{\text{ext}}$ are applied in 
        tetrahedral direction. In this case, a tetrahedral 
        quadruple-junction forms and remains stable.
	\label{fig:figS1}
    }
\end{figure}

As mentioned in the main text, higher order junctions tend to 
split up into Y-junctions in the simulations with applied mechanical
force. Fig.\ \ref{fig:figS1}a,b) shows an example of such 
an evolution.
\fs{
The following movies are provided along with this article to illustrate
this further:
\begin{itemize}
    \item {\tt 4fold.mp4}: Time evolution after applying four coplanar
       forces with angles $90^0$ (Fig. \ref{fig:figS1}b).
    \item {\tt 4fold\_twisted.mp4}: Time evolution after applying four 
       forces with angles $90^0$, which are twisted with respect to each other.
    \item {\tt tetrahedral.mp4}: Time evolution after applying four 
       forces with with angles corresponding to a tetrahedral junction. 
       Under the influence of external forces, the tetrahedral junction 
       is stable
       (Fig. \ref{fig:figS1}c).
\end{itemize}
}

\subsubsection{Force-free structures}

\fs{As explained in the main article, branched structures and 
thin linear structures undergo a transformation to a stomatocyte structure 
if the stabilizing forces are released at fixed reduced volume $\nu$. 
This is shown in the following movies:
\begin{itemize}
   \item {\tt f0\_branch\_nu14.mp4}: Time evolution after releasing
    the forces from an initially branched structure with a
    Y-junction while keeping $\nu = 0.14$ fixed. The transition  
    sets in instantaneously.
     \item {\tt f0\_branch\_nu19.mp4}: Time evolution after releasing
    the forces from an initially branched structure with a
    Y-junction while keeping $\nu = 0.19$ fixed. The transition
    sets in instantaneously.
    \item {\tt f0\_linear\_nu14.mp4}: Time evolution after releasing
    the forces from an initially linear structure while keeping
    $\nu = 0.14$ fixed. The transition sets in after an activation time.
     \item {\tt f0\_linear\_nu19.mp4}: Time evolution after releasing
     the forces from an initially linear structure while keeping
    $\nu = 0.19$ fixed. Within the simulation time, no transition
    takes place.
\end{itemize}
}

\begin{figure}
         \centering
         \includegraphics[width=1.05\linewidth]
         {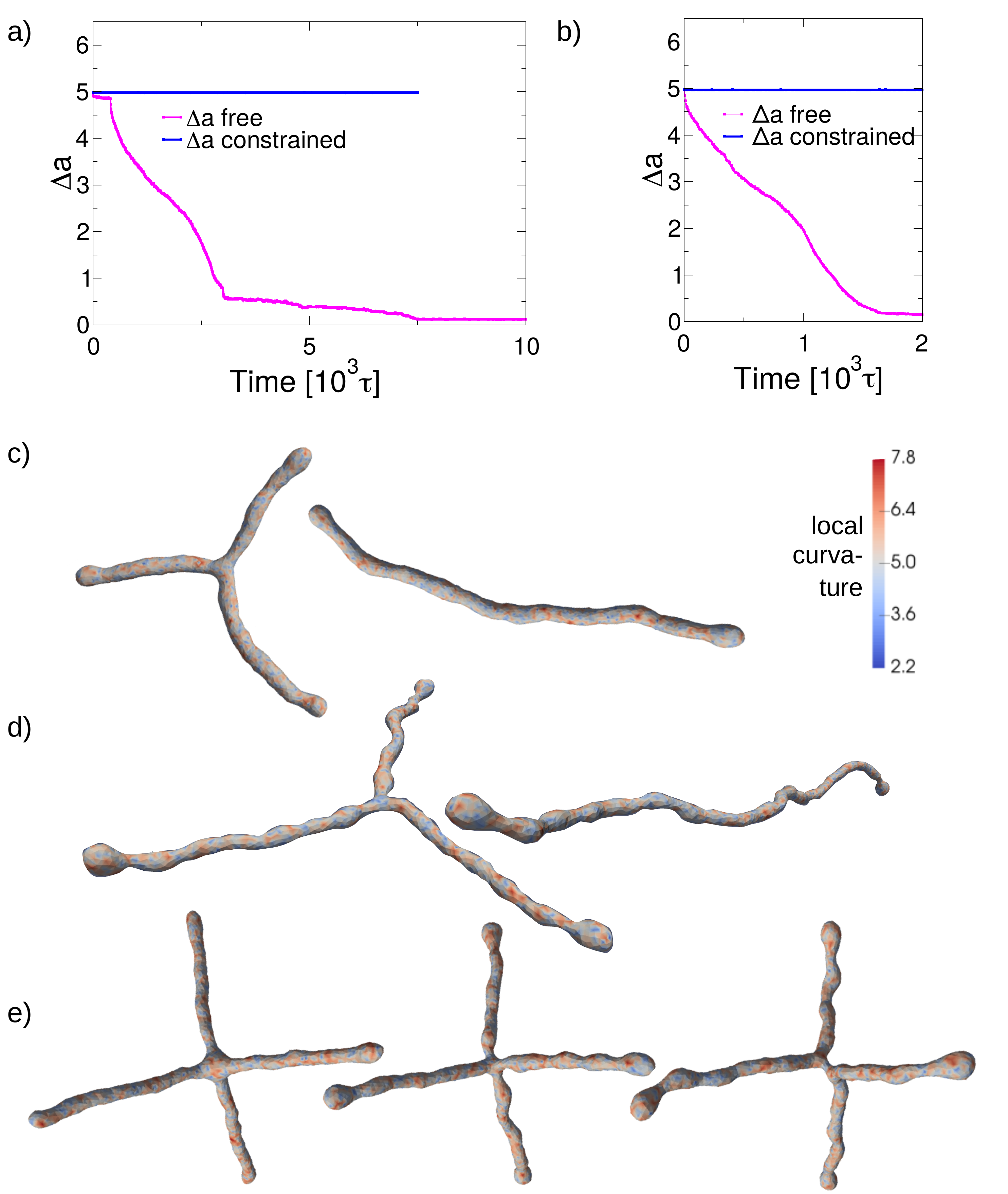}
     \caption{
     (a,b) Time evolution of the rescaled area difference $\Delta a$
     after releasing the force on a force-stabilized \fs{linear} (a) and
     branched (b) structure at fixed reduced volume $\nu=0.14$
     (pink curve). If $\Delta a$ is kept fixed (blue curve), the
     tubular-stomatocyte transformation is suppressed, and
     the linear and branched structures persist.
     (c) Examples of force-free metastable \fs{linear} and branched
     structures for fixed $\nu=0.27$ and fixed $\Delta a$.
     \fs{(d) Examples of stable structures for fixed $\Delta a$
     and unconstrained $\nu$
     (e) Time evolution of an initially force-stabilized structure
      containing a tetrahedral junction at fixed $\nu = 0.2$ and
     $\Delta a$. The tetrahedral junction splits up into 
     two Y-junctions.
     The color coding in c-e) indicates distribution of local
     curvature on the surfaces.
     }
      \label{fig:figS2}
      }
 \end{figure}

\fs{Constraining the average curvature $\Delta a$ in addition to the
reduced volume $\nu$ 
stabilizes the tubular structures, and both linear and branched 
tubular structures remain stable (see Fig. \ref{fig:figS2} a-d). 
Tetrahedral junctions however do not survive, they split up into 
two Y-junctions. This is shown in Fig. \ref{fig:figS2}e) and 
in the movie   {\tt f0\_tetrajedral\_nu20fixDa.mp4}. 
}

\subsection{Curvature energy balance of tubular structures at fixed
$\Delta a$}

We wish to estimate the energy difference between pure tubes and
branched structures with three arms under the condition that the
area difference, $\Delta a \propto I_K := \int \text{d}A \: K$ is fixed.

First we estimate the local excess energies of caps and junctions.

Our reference system is a pure cylinder (no caps) with area $A$ and
tube radius $R$, which has the curvature energy
$E_{_{\text{CV}}}=A/2R^2$ and the integrated curvature
$I_K = A/R$. Compared to this reference, each spherical cap 
contributes an excess free energy $\Delta E_{\text{cap}}= 3 \pi
\kappa$ and an excess integrated curvature
$\Delta I_{K,\text{cap}} = 2 \pi R$. Correspondingly,
each junction contributes an excess free energy
$\Delta E_{\text{junction}}$ and an excess integrated
curvature $\Delta I_{K,\text{junction}}$, which are both
unknown. However, we can approximately relate them to each
other by assuming that the junction can be described as 
a section with area $A_{\text{junction}}$ with reduced
mean total curvature $\alpha/R$, \mbox{$\alpha < 1$.}
This implies
\begin{eqnarray}
\label{eq:corr1}
\Delta I_{K,\text{junction}} 
&\approx& (\alpha - 1)\frac{A_{\text{junction}}}{R},
\\
\label{eq:corr2}
\Delta E_{\text{junction}} 
&\approx &
(\alpha^2 - 1) \kappa \frac{A_{\text{junction}}}{2R^2}
\end{eqnarray}
\fs{From Fig.\ 2c in the main article},
we know $\Delta E_{\text{junction}} \approx -280$,
roughly independent of the tube radius, hence
$\alpha < 1$ and $A_{\text{junction}}/R^2 
  =: a_{\text{junction}} \approx $const.

The curvature energy and the integrated curvature of
tubes and three-arm structures with radius $R$ are estimated as
\begin{eqnarray}
\label{eq:ectube}
\frac{1}{\kappa} E_{_{\text{CV}},\text{tube}} &=&
\frac{A}{2 R^2} + 6 \pi \\
\label{eq:ecbranch}
\frac{1}{\kappa} E_{_{\text{CV}},\text{branch}} &=&
\frac{A}{2 R^2} + 9 \pi + \frac{\Delta E_{\text{junction}}}{\kappa}
\\
\label{eq:iktube}
I_{K,\text{tube}} &=&
\frac{A}{R} + 4 \pi R \\
\label{eq:ikbranch}
I_{K,\text{branch}} &=&
\frac{A}{R} + 6 \pi R 
 + \Delta I_{K,\text{junction}} 
\end{eqnarray}

For given fixed $R$, the curvature energy of branched structures
and pure tubes thus differs by 
\begin{equation}
E_{_{\text{CV}},\text{branch}}-E_{_{\text{CV}},\text{tube}}
= 3 \pi \kappa + \Delta E_{\text{junction}}
\approx -90
\end{equation}
which is negative as discussed in the main article.

However, if $I_K$ is kept fixed, then both the radii
of the pure tube, $R_{\text{tube}}$ and of the branched
structure, $R_{\text{branch}}$, change with respect to
the value $\bar{R}$ in the reference cylinder.
Denoting $\bar{c}=1/\bar{R}$ and
$\Delta c = 1/R - 1/\bar{R}$, Eq.\ (\ref{eq:iktube})
yields
\begin{equation}
\Delta c_{\text{linear}} = - \frac{4 \pi}{A} \: R_{\text{linear}}
\end{equation}
for \fs{linear structures}, and Eq.\ (\ref{eq:ikbranch})
\begin{equation}
\Delta c_{\text{branch}} = 
- \frac{1}{A} \Big( 6 \pi \: R_{\text{branch}}
   +  \Delta I_{K,\text{branch}} \Big) \:
\end{equation}
for branched three-arm structures.
Inserting this in Eqs.\ (\ref{eq:ectube}), (\ref{eq:ecbranch}), we
obtain
\begin{eqnarray}
\nonumber
\frac{1}{\kappa} E_{_{\text{CV}},\text{linear}} &=&
\frac{A}{2} (\bar{c}+\Delta c)^2 + 6 \pi 
\\ &=&
\frac{A}{2} \bar{c}^2 + 2 \pi + {\cal O}(\frac{1}{A})
\\
\nonumber
\frac{1}{\kappa} E_{_{\text{CV}},\text{branch}} &=&
\frac{A}{2} (\bar{c}+\Delta c)^2 
+ 9 \pi  + \frac{\Delta E_{\text{junction}}}{\kappa}
\\ &=&
\frac{A}{2} \bar{c}^2 + 3 \pi 
+ \frac{a_{\text{junction}}}{2} (\alpha - 1)^2
\nonumber
\\ &&
+ \: {\cal O}(\frac{1}{A})
\end{eqnarray}
\begin{figure}[h!]
		\centering
		\includegraphics[width=1.0\linewidth]
        {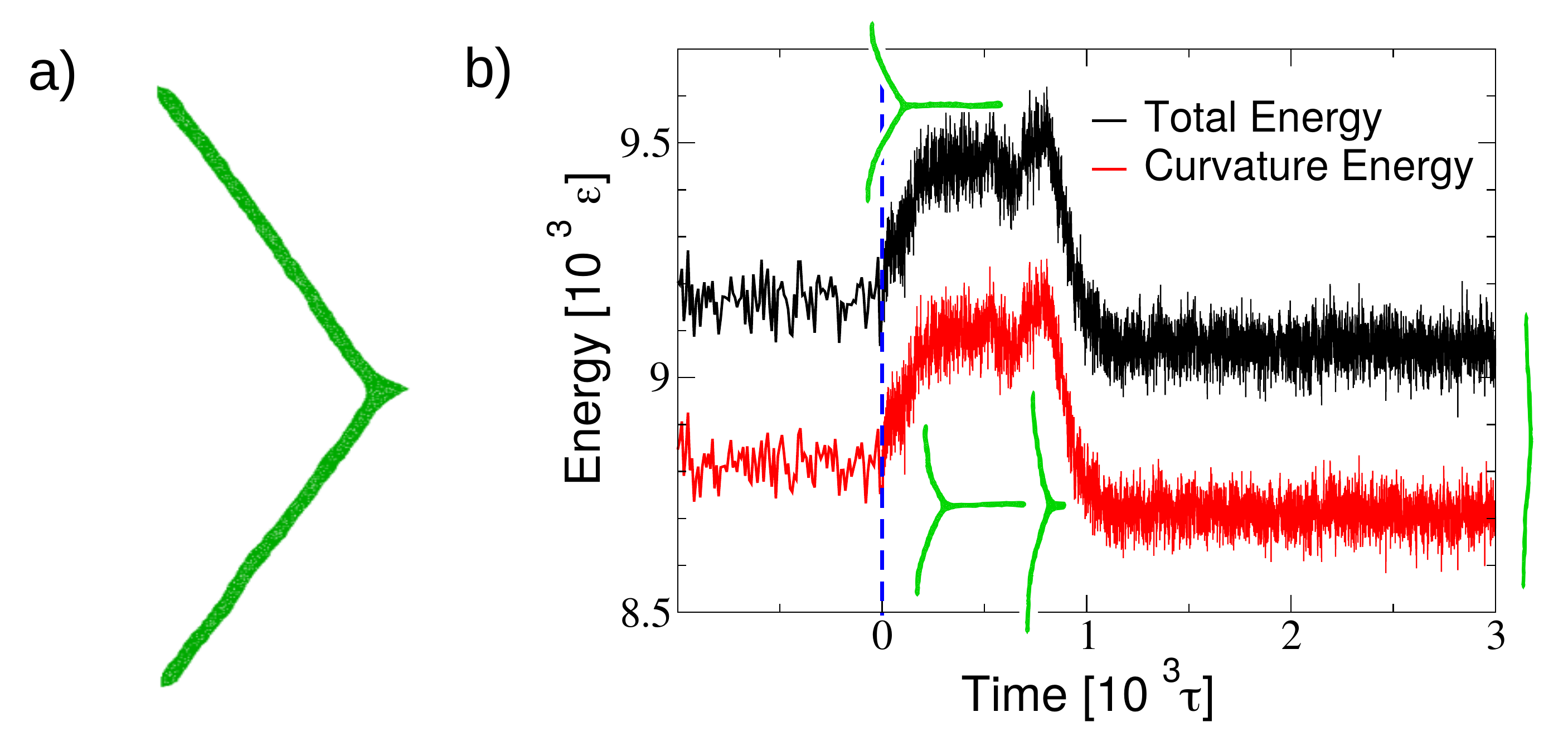}
	\caption{
    Transition between \fs{linear} and branched structures
    (a) Metastable configuration of a \fs{linear tubular structure} 
        when applying a force to pull a branch out from the tubular 
        structure, using the same external forces as needed to 
        stabilize branches. The tube deforms, but does \fs{no branch
        emerges}. ($F_{\text{ext}}=90$, no constraints).
    (b) Time evolution after releasing the force on one tube 
        in a branched structure and pulling the other two
        tubes with in opposite directions ($F_{\text{ext}}=70$).
        $\Delta a$ and $\nu$ are constrained to fixed values.
        The force is released at the time $t=0$. It
        retracts and eventually disappears at the time
        $t \approx 2\times 10^3 \tau$. 
        During this process, the curvature energy rises,
        suggesting the presence of a free energy barrier.
	\label{fig:figS3}
    }
\end{figure}

where we have used 
$R_{\text{\fs{linear}/branch}}\bar{c} = 1 + {\cal O}(\frac{1}{A})$
and Eqs.\ (\ref{eq:corr1},\ref{eq:corr2}).
Hence the difference between \fs{the energy of branched and
linear structures} for fixed area difference is approximated by
\begin{equation}
E_{_{\text{CV}},\text{branch}}-E_{_{\text{CV}},\text{linear}}
\approx \pi \kappa 
+ \frac{\kappa}{2} \: a_{\text{junction}} \: (\alpha-1)^2
\end{equation}
which is now always positive.
In terms of $\Delta E_{_{\text CV},\text{junction}}$,
this expression can be rewritten as
\begin{equation}
E_{_{\text{CV}},\text{branch}}-E_{_{\text{CV}},\text{linear}}
= \pi \kappa  + | \varepsilon \cdot \Delta
E_{_{\text{CV}},\text{junction}} |.
\end{equation}
with $\varepsilon = (1-\alpha)/(1+\alpha)$.

\fs{We should note that the final result does not depend on the sign of $\Delta E_{\text{junction}}$. The underlying reason is that it is always more favorable to distribute a given integrated curvature as homogeneously as possible on a fixed surface than to allow for local variations.}


\medskip

\subsection{Transitions between pure \fs{linear} and branched structures}

To investigate transitions between \fs{linear} and branched structures,
we have carried out two types of simulations.  

First, starting from a force-stabilized \fs{linear tubular} structure, we 
apply forces $F_{\text{ext}}$ at the tube ends and \gj{an} additional
point in the middle with angles $120^o$. The tube deforms and assumes 
a V-shape, but no branch forms out (Fig.\ \ref{fig:figS3}a). In order 
to pull an additional branch out, the additional force must be increased 
by $\Delta F_{\text{ext}} \approx 3-4$, which is approximately 4-6\%
higher than the force needed for stabilization (at $F_{\text{ext}} =
60-1000$).

Second starting from a force-stabilized branched structure, we release
the force on one branch and replace the forces on the other two ends
by two forces $F_{\text{ext}}$ in opposite direction. This is done
at fixed $\Delta a$ and $\nu$. The loose branch retracts and
eventually vanishes. This is associated with an increase of
curvature energy
(Fig.\ \ref{fig:figS3}b). A similar barrier is observed if only
$\nu$ is kept fixed. If both $\nu$ and $\Delta a$ are unconstrained,
no energy barrier is observed, instead the curvature energy even
drops during the transition.

\end{document}